# A 2D IR Study of Isotope-Edited Variants of the Elastin-like GVGVPGVG Peptide and the Size Dependent Behavior of (VPGVG)$_n$

Joshua A. Lessing

**Abstract:**

This study uses two-dimensional infrared (2D IR) spectroscopy in conjunction with isotope labeling and spectral modeling from molecular dynamics simulations to identify the dominant turn conformations that exist in equilibrium ensembles of the (VPGVG)$_n$ family of intrinsically disordered elastin-like peptides. Numerous models have been proposed to explain the origins of elastin's elasticity and its counterintuitive ability to become structured upon heating. However, the structure of elastin remains unassigned because none of the techniques currently used to study highly disordered amino acid sequences have a time resolution that is faster than the lifetime of a transient conformation in a disordered sequence. Because these conformations exchange on time scales longer than the 5-6 ps required for 2D IR measurements, isotope-edited 2D IR spectroscopy was chosen to study this family of peptides. This study first examined the small GVGVPGVG peptide at a series of temperatures and salt concentrations to assign its dominant turn conformations and to determine what variables alter the populations of these conformations. These data were then used to identify the dominant turn conformations that exist in larger elastin-like peptides including (GVGVP)$_{251}$. The results indicate that the (VPGVG)$_n$ family of elastin-like peptides contain a high population of both an irregular turn structure with 2 peptide-peptide hydrogen bonds to the proline amide C=O group and a conventional turn structure with 1 peptide-peptide hydrogen bond to the proline amide C=O group. These turn structures are durable, showing a significant population of closed turns at all temperatures and salt concentrations studied.

# 1 Introduction

Intrinsically disordered sequences (IDSs) are amino acid sequences that do not form stable secondary or tertiary structures in solution under physiological conditions. Instead, IDSs exist as heterogeneous ensembles of conformations that may interconvert on timescales from picoseconds to microseconds. Once considered rare and ineffectual, IDSs have been shown to play crucial roles in biological systems.[1-3] Bioinformatics-based predictions estimate that IDSs are common in proteins responsible for physiologic and pathologic regulation, recognition, and signaling.[4-5] Intrinsically disordered regions (IDRs) and intrinsically disordered proteins (IDPs) are predicted to be heavily expressed in numerous diseases including cancer, diabetes, Parkinson's, Alzheimer's and cardiovascular disease.[4-7] As a result, developing the means to characterize heterogeneous ensembles of IDSs is a necessary step for the advancement of human health and disease prevention.

Detailed structural characterization of samples containing high percentages of IDSs is difficult using X-ray crystallography, fluorescence, NMR, ESR and SAXS.[8-11] Nevertheless, these techniques have been employed to gain valuable insight into the properties of IDSs. For example, IDSs of ≥ 30 amino acids and 10 to 29 amino acids are observed in ~10% and ~40%, respectively, of the X-ray crystal structures in the PDB.[11] The recent identification of patterns in these sequences via the use of bioinformatics-based methods has aided in the creation of algorithms for predicting

IDSs. In solution NMR experiments, the deviations of the chemical shifts and NOE values from their expected random coil values have been used to detect transient secondary structures.[12-14] Unfortunately, none of the techniques currently used to study IDSs have time resolutions that are faster than the lifetime of a transient conformation in a highly disordered sequence. As a result, averaged measurements are used for structure determination, which diminishes the ability to resolve structural heterogeneity in ensembles. However, isotope-edited two-dimensional infrared spectroscopy (2D IR) is a promising technique for the study of IDSs because of its ability to measure protein secondary structure with picosecond time resolution.[15-21] Because the conformations of IDSs exchange on time scales longer than the 5-6 ps required for the 2D IR measurement, 2D IR can be used to distinguish between transient structures in a heterogeneous ensemble.

In this report, we use 2D IR spectroscopy in conjunction with spectral modeling from molecular dynamics simulations to identify the dominant structures that exist in the equilibrium ensemble of an elastin-like peptide. Elastin, which is a durable elastomeric IDP found in the extracellular matrix of animal connective tissue, has been the subject of extensive investigation owing to its distinctive physical properties.[22-32] For example, while the elastin protein exists in an extended state at low temperatures, the protein contracts and aggregates to form a coacervate phase upon heating, salt addition and/or increasing pH.[33-39]

Numerous models have been proposed to explain the origins of elastin's elasticity and its counterintuitive ability to become structured upon heating, known as an inverse temperature transition (ITT).[40] These models include the phenomenological single-phase classical rubber[41-42] and liquid drop models[43] as well as the structure-based oiled coil,[44] β-spiral,[45] cis-trans isomerization,[46] and three phase models.[47] However, because the structure of elastin remains unassigned, the molecular mechanism of elastin's phase transition and the origins of its elasticity are still up for debate.

The limited information about the molecular structure of elastin is due to both the fibrous structure of the natural protein and the high mobility of its peptide backbone. As a result, the mature elastin fiber cannot be crystalized and is therefore unsuitable for X-ray diffraction. In addition, the fiber is highly insoluble in water and therefore unsuitable for solution NMR studies. However, solid-state NMR experiments have recently been performed, providing insight into the structure of dry and semi-hydrated elastin fibers.[48-49] Nevertheless, a solid-state NMR-based structure of elastin under more native hydrated conditions has remained elusive due to the increase in the mobility of elastin peptide chains upon the addition of water.

To work around these difficulties, many researchers have studied water-soluble tropoelastin and peptides consisting of repeating units of amino acid sequences found in the natural protein; these sequences are termed elastin-like peptides (ELPs). With the exception of small cyclic ELPs,[50-51] these samples cannot be crystalized and have fast conformational motions that are difficult to resolve with NMR. Because 2D IR spectroscopy can provide structural information on fast time scales and can interrogate non-crystalline samples, it is an ideal tool for studying elastin. Therefore, we use this technique to analyze an ELP in this work.

## 2 Experimental Overview

The $^{13}$CO and $^{13}$C$^{18}$O isotopologues synthesized for this work are shown in Figure 1. This library includes the unlabeled or "wild type" (WT) peptide, 5 single labeled peptides and 3 multiply labeled peptides. The names assigned to each amide site and listed from the N to the C terminus are G3†, V4†, G5†, V1, P2, G3, and V4.

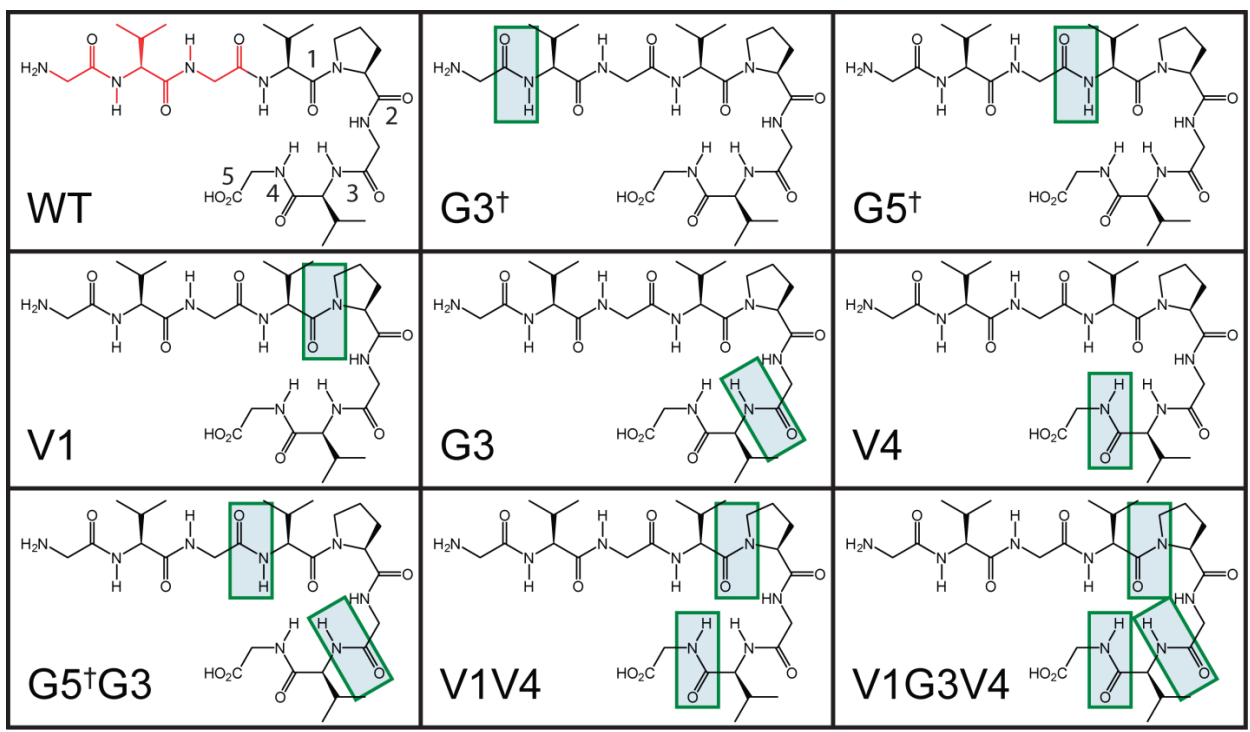

**Figure 1.** Library of peptides used to analyze the GVGn1 structure. The boxes indicate the locations of the isotope-labeled amide sites. All labeled amide sites were labeled with $^{13}C^{18}O$, except for the G3 site in V1G3V4, which was labeled with $^{13}CO$ for increased peak resolution. The monomer unit VPGVG is drawn in black in the WT structure with amide sites numbered to demonstrate the naming convention used in this work.

This naming scheme reflects the fact that VPGVG is a monomer unit in the elastin biopolymer. As a result, the N terminal sequence G3†V4†G5† is numbered in accordance with their position in the previous monomer unit, which is distinguished with a †. This scheme assigns the name of an amide unit based on the amino acid to its N terminus. For example, an isotope label on the amide unit between G3 and V4 is referred to as the G3 label.

Five single $^{13}C^{18}O$ labeled peptides were synthesized: the turn labels V1, G3 and V4; the mid-strand label G5†; and the N-terminal label G3†. Isotope labeling was performed to decouple the vibrations of the labeled amide site from those of its neighbors providing a local probe of peptide structure. Studying the line shape of the labeled peak elucidates local information on hydrogen bonding, structural heterogeneity and the rate at which these heterogeneous environments interconvert. In addition, evaluating the frequency-dependent changes in the main amide I′ band and cross peaks generated upon labeling provides clues about the contribution of the sites to the secondary structure of the peptide.

The dual $^{13}C^{18}O$ labels G5†G3 and V1V4 and the triple labeled V1G3V4 (with $^{13}C^{18}O$ on the V1 and V4 sites and $^{13}CO$ on the G3 site) were synthesized to measure peptide secondary structure, analogous to the TT labeled Trpzip2 peptide in the work of Smith et al.[52] In the case of Trpzip2, the coupling of the T3 and T10 isotope labels across a semi-ordered β-sheet structure generated a measurable shift in the site frequency of the TT peak relative to the single label peak for the T10 peptide. The same approach was used for the multiply labeled peptides by comparing the FTIR and 2D IR spectra for each multiple label to their corresponding single label analogs. The labels in the G5†G3, V1V4, and V1G3V4 peptides are located at positions around the turn and can report on changes in the structure of the turn.

These labels are first presented at 10 °C to provide a description of the average structure of the peptide. Next, temperature and salt dependent data will be presented to demonstrate the changes that occur in the structure of the turn as the peptide is driven through the ITT. Finally, data for $GVG(VPGVG)_n$ where n = 1-6 and $(GVGVP)_{251}$ will be presented to determine if the length of the peptide effects the turn conformation.

## 3 Results and Discussion

### FTIR and 2D IR spectra

The FTIR spectra for the desalted GVGn1 isotopologues at 10 °C in pH = 1.0 DCl in $D_2O$ with 150 mmol of $KD_2PO_4:K_2DPO_4$ (1:1 by mol) are presented in Figure 2. The amide I′ spectrum of GVGn1 contains several notable peak features, which will be briefly reviewed in this section.

The peak at $\omega$ = 1614 cm$^{-1}$ in the FTIR spectra of WT, G3$^{\dagger}$, G5$^{\dagger}$, G3, V4 and G5$^{\dagger}$G3 is attributed to the amide I′ vibration of proline. The peaks at $\omega$ = 1640 and 1670 cm$^{-1}$ are assigned to the $v_{\perp}$ and $v_{\parallel}$ antiparallel β-sheet modes.[53-54] The peak at $\omega$ = 1660 cm$^{-1}$ is assigned to random coil vibrations because GVGn1 is an eight residue peptide with a central proline and is therefore unable to form an α-helix. Finally, the non-zero offset on the blue side of the spectrum is the result of the COOD stretching mode that is centered at ~1725 cm$^{-1}$. Isotope-labeled peaks appear on the red side of the spectrum with peak maxima at frequencies ranging from 1554.4 cm$^{-1}$ for the $^{13}$C$^{18}$O-labeled V1 site in the V1 peptide to 1610.3 cm$^{-1}$ for the $^{13}$CO-labeled G3 site in the V1G3V4 peptide. The combination of the $^{13}$CO, $^{13}$C$^{18}$O, and proline redshifts of 41, 60, and 27 cm$^{-1}$, respectively, can provide a high degree of site specific resolution in a single spectra as evidenced by the V1G3V4 peptide.

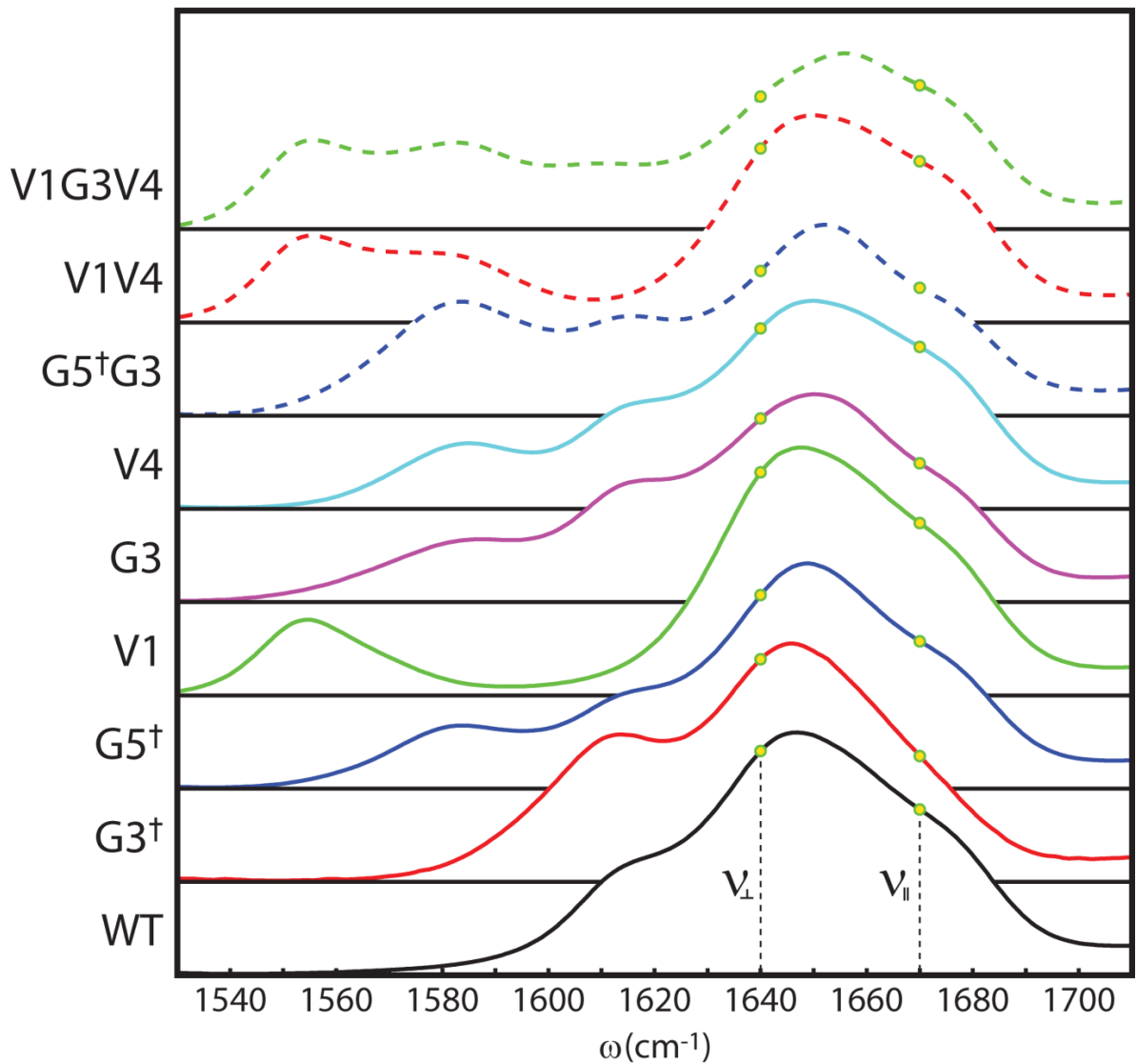

**Figure 2.** FTIR spectra for the GVGn1 peptide library collected at 10 °C on desalted peptides in pH = 1.0 DCl in $D_2O$ with 150 mmol of $KD_2PO_4$:$K_2DPO_4$ (1:1). The green and yellow circles are placed at 1635 and 1675 cm$^{-1}$ to indicate the location of the β-sheet $v_\perp$ and $v_\parallel$ modes, respectively.

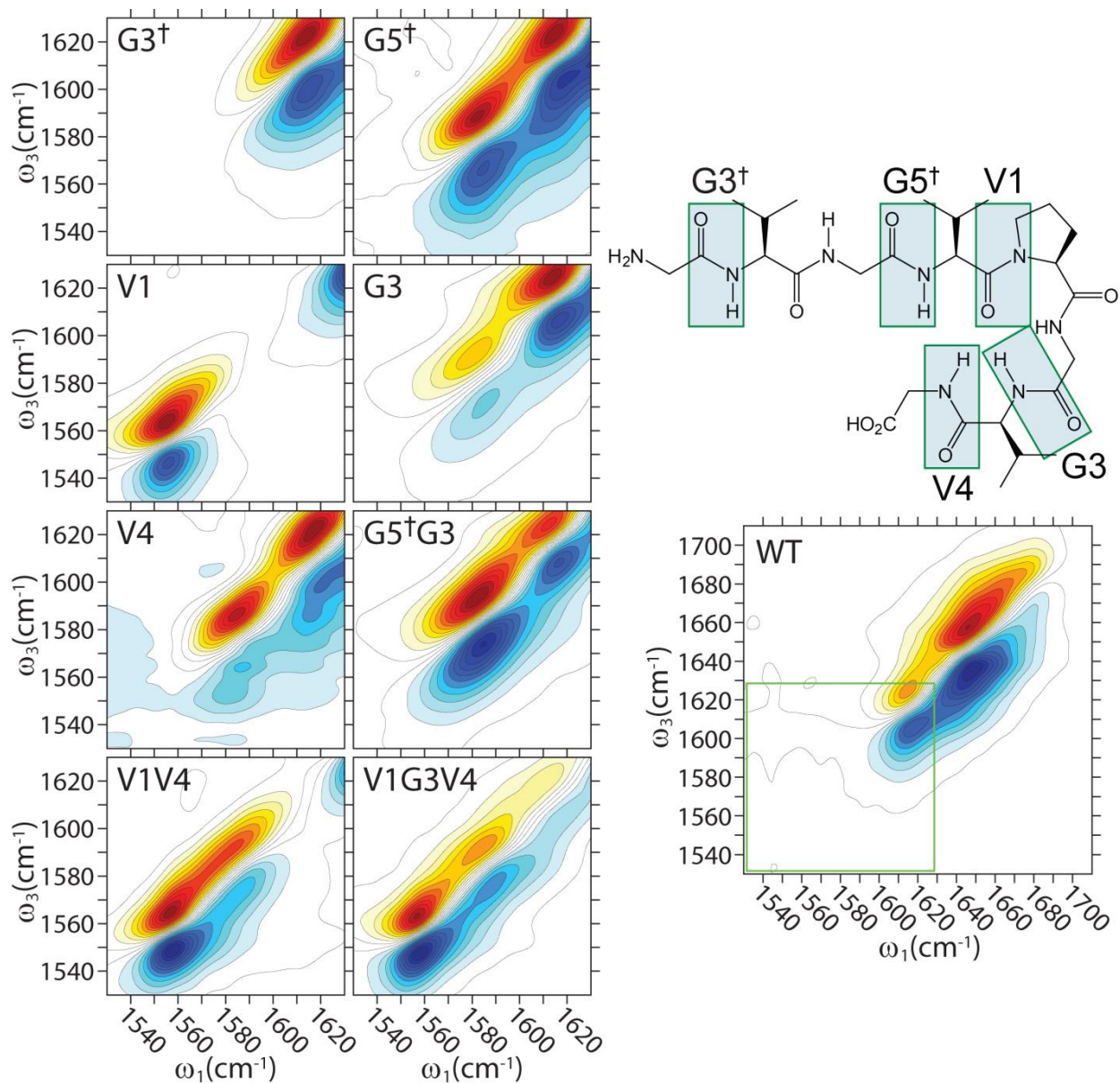

**Figure 3.** 2D IR spectra for the library of GVGn1 peptides used to analyze the peptide structure. Spectra were collected in ZZYY polarization at 10 °C and $\tau_2 = 150$ fs on desalted peptides in pH = 1.0 DCl in D$_2$O with 150 mmol of KD$_2$PO$_4$:K$_2$DPO$_4$ (1:1). Due to the desalting of the peptides, these 150 mmol spectra contain line shapes that are similar to the spectra collected for peptides prepared without a desalting step and dissolved in pH = 1.0 DCl in D$_2$O with no additional salt added.

The 2D IR spectra for the desalted GVGn1 isotopologues at 10 °C with a waiting time of $\tau_2 = 150$ fs in pH = 1.0 DCl in D$_2$O with 150 mmol of KD$_2$PO$_4$:K$_2$DPO$_4$ (1:1) are presented in Figure 3. Due to the non-linear scaling of intensity in 2D IR surfaces, peaks are better resolved in 2D IR than in FTIR spectra, making the identification of peak features from a 2D IR surface a robust metric for characterization. With the exception of G3$^\dagger$, the 2D IR surfaces for all of the peptides contain well-resolved isotope peaks. As a result, these data sets can be easily fit with Gaussian peaks and interpreted using basic line shape metrics, such as center frequency ($\omega$), on-diagonal FWHM ($\Delta$), off-diagonal FWHM ($\Gamma$), peak rotation and ellipticity (E = $(\Delta^2 - \Gamma^2)/(\Delta^2 + \Gamma^2)$). Fitting the 2D IR surfaces gave $^{13}C^{18}O$ isotope peak frequencies of G5$^\dagger$: $\omega$ = 1582.0 cm$^{-1}$, V1: $\omega$ = 1554.4 and 1572.2 cm$^{-1}$, G3: $\omega$ = 1581.0 cm$^{-1}$, and V4: $\omega$ = 1584.1 cm$^{-1}$. The isotope peak frequency of the G3$^\dagger$ peptide could not be extracted from the 2D IR spectrum because its resonance is nearly degenerate with that of unlabeled proline. As a result, FTIR difference spectra were used to assign the isotope peak frequency, yielding an isotope peak of G3$^\dagger$: $\omega$ = 1604.4 cm$^{-1}$ (Figure 4). The values for $\omega$, $\Delta$, $\Gamma$, and E are presented for all labeled sites in Table 1.

|  | $\omega$ | $\Delta$ | $\Gamma$ | E |
|---|---|---|---|---|
| G3$^\dagger$ | 1604.4 | | | |
| G5$^\dagger$ | 1582.0 | 25.5 | 9.6 | 0.75 |
| V1 (2/0) | 1554.4 | 19.3 | 9.3 | 0.62 |
| V1 (1/0) | 1572.2 | 16.3 | 10.0 | 0.46 |
| G3 | 1581.0 | 34.6 | 10.4 | 0.84 |
| V4 | 1584.1 | 24.4 | 7.6 | 0.82 |

**Table 1.** Center frequency ($\omega$), on diagonal FWHM ($\Delta$), off diagonal FWHM ($\Gamma$), and ellipticity (E) for the isotope-labeled sites of GVGn1. The values for $\omega$, $\Delta$, and $\Gamma$ are presented here in cm$^{-1}$, and the value for the ellipticity is given by E = $(\Delta^2 - \Gamma^2)/(\Delta^2 + \Gamma^2)$. Peak metrics for all compounds with the exception of G3$^\dagger$ were measured by fitting their 2D IR spectra. Because the G3$^\dagger$ isotope peak is partially overlapped with the proline resonance, it was not possible to accurately fit its 2D IR surface. As a result, the value for $\omega$ was measured from the [G3$^\dagger$ - WT] FTIR difference spectrum shown in Figure 4.

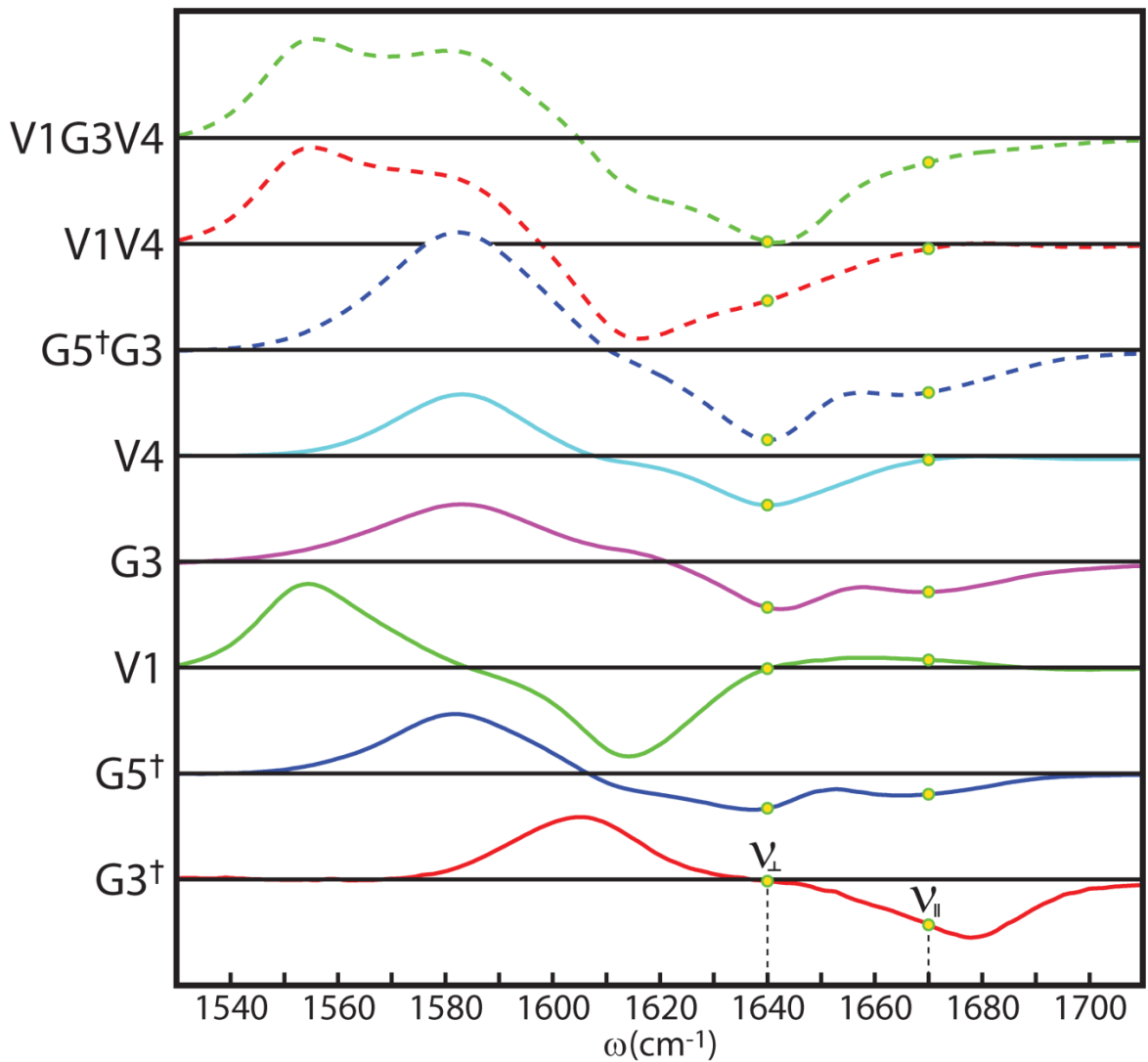

**Figure 4:** [(Isotope Label) – WT] FTIR difference spectrum for the GVGn1 peptide library collected at 10 °C on desalted peptides in pH = 1.0 DCl in $D_2O$ with 150 mmol of $KD_2PO_4$.

## Single Site Analysis: V1

Peak frequency values provide insight into the local environment surrounding each labeled site. Based on the peak frequency of the proline resonance, Lessing et al.[54] established that the GVGn1 peptide contains a high population of 1/0 and 2/0 hydrogen bond turns, where here the first digit corresponds to the number of hydrogen bonds the proline CO accepts from other amide groups in the peptide and the second digit corresponds to the number of hydrogen bonds the proline CO accepts from water. This assertion is confirmed here by the presence of an asymmetric V1 peak line shape (Figure 5).

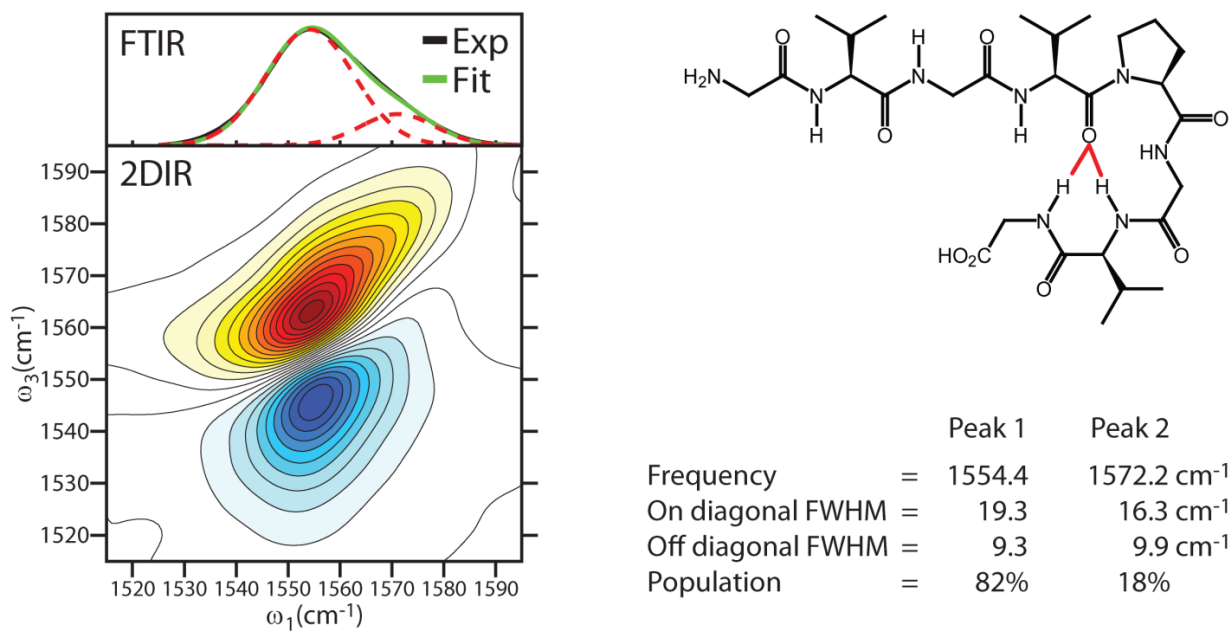

**Figure 5.** Experimental V1 FTIR and ZZYY 2D IR spectrum collected at 10 °C on a desalted peptide at $\tau_2$ = 150 fs in pH = 1.0 DCl in $D_2O$ with 150 mmol of $KD_2PO_4$. The FTIR spectrum was fit with 2 Gaussian peaks, giving populations for the 2/0 and 1/0 turns of 82% and 18%. The upper right hand corner of the figure contains a schematic representation of the GVGn1 peptide in a 2/0 turn with hydrogen bonds depicted as red lines.

The asymmetric peak is due to the overlap of the 1/0 and 2/0 peaks at 1572.2 and 1554.4 cm$^{-1}$, respectively. The frequency difference of $\Delta$ = 17.8 cm$^{-1}$ is more similar to the 16 cm$^{-1}$ redshift per hydrogen bond found for the K8 label of Trpzip2 in Smith et al.[52] than the 13.8 cm$^{-1}$ redshift per hydrogen bond found for the proline site in Lessing et al.[54] Based on the fitting of the 2D IR spectrum, the V1 peak was found to have a $\Delta$ = 19.3 cm$^{-1}$, $\Gamma$ = 9.3 cm$^{-1}$, and E = 0.62 for the 2/0 peak and $\Delta$ = 16.3 cm$^{-1}$, $\Gamma$ = 9.9 cm$^{-1}$, and E = 0.46 for the 1/0 peak. Based on the 3 cm$^{-1}$ larger $\Delta$, the 2/0 structure contains a more heterogeneous distribution of local environments. In addition, because the 2/0 peak has a 1.2 cm$^{-1}$ smaller $\Gamma$ and a 35% greater ellipticity relative to the 1/0 peak, the ensemble of 2/0 turns likely exchanges less rapidly than the 1/0 ensemble. Fitting to the V1 FTIR isotope peak was used to assign the relative populations of the 1/0 and 2/0 turns, yielding 18% and 82%, respectively. These values disagree with the results shown in Lessing et al.[54] where the populations for the 1/0 and 2/0 turns were reported to be 55-60% and ~25%, respectively. Nevertheless, both predictions show high populations of closed turns. In contrast, the MD simulations predict an open turn with an 11.5% and 4.9-5.9% population for the 1/0 and 2/0 turn, respectively.[54] As demonstrated previously[54], only peptide conformations with closed turns reproduce the pronounced proline peak shift in the GVGn1 spectrum. As a result, the values presented here and in Lessing et al.[54], which predict high populations of closed turns, are an improvement over MD-based results. These discrepancies also demonstrate the value of isotope labeling, which was used here to fully isolate peak features: the combination of the proline and $^{13}C^{18}O$ redshifts changed the V1 site frequency by $\Delta\omega = -87$ cm$^{-1}$, allowing for elucidation of the results presented in this work.

The 2D IR data presented above were collected at a waiting time of $\tau_2 = 150$ fs, which is fast relative to the time scale of molecular motion. The result is a 2D IR surface that presents a nearly static picture of the conformational ensemble. By studying the isotope-labeled peaks as a function of waiting time, it is possible to track the configurational changes that occur for this ensemble providing insight into the molecular dynamics of the peptide.[55] These isolated peaks transition from diagonally elongated to symmetric as a function of waiting time, which indicates that the structures of the local environments, and thus the amide I′ transition frequencies of the sites, evolve. Information about the local solvent exposure, conformational flexibility, and chemical exchange is encoded in these waiting-time-dependent spectral changes. The value of $\Delta$ for an isotope peak is the result of the frequency for that transition being modulated by the distribution of heterogeneous local environments that exist for that labeled amide site in the ensemble. Transitions that were initially at a frequency of $\omega_1$ evolve as a function of waiting time to a new frequency $\omega_3$ due to changes in the local environment, resulting in the generation of two off diagonal peaks at ($\omega_1$, $\omega_3$) and ($\omega_3$, $\omega_1$). In the case of a rearrangement of the local solvent environment or a transient conformational fluctuation of the peptide, these cross peaks generate a symmetric line shape with a rotation of the tilt of the on diagonal peak maximum. In the case of chemical exchange, these cross peaks can generate a distinct feature in the off diagonal region of the spectrum.

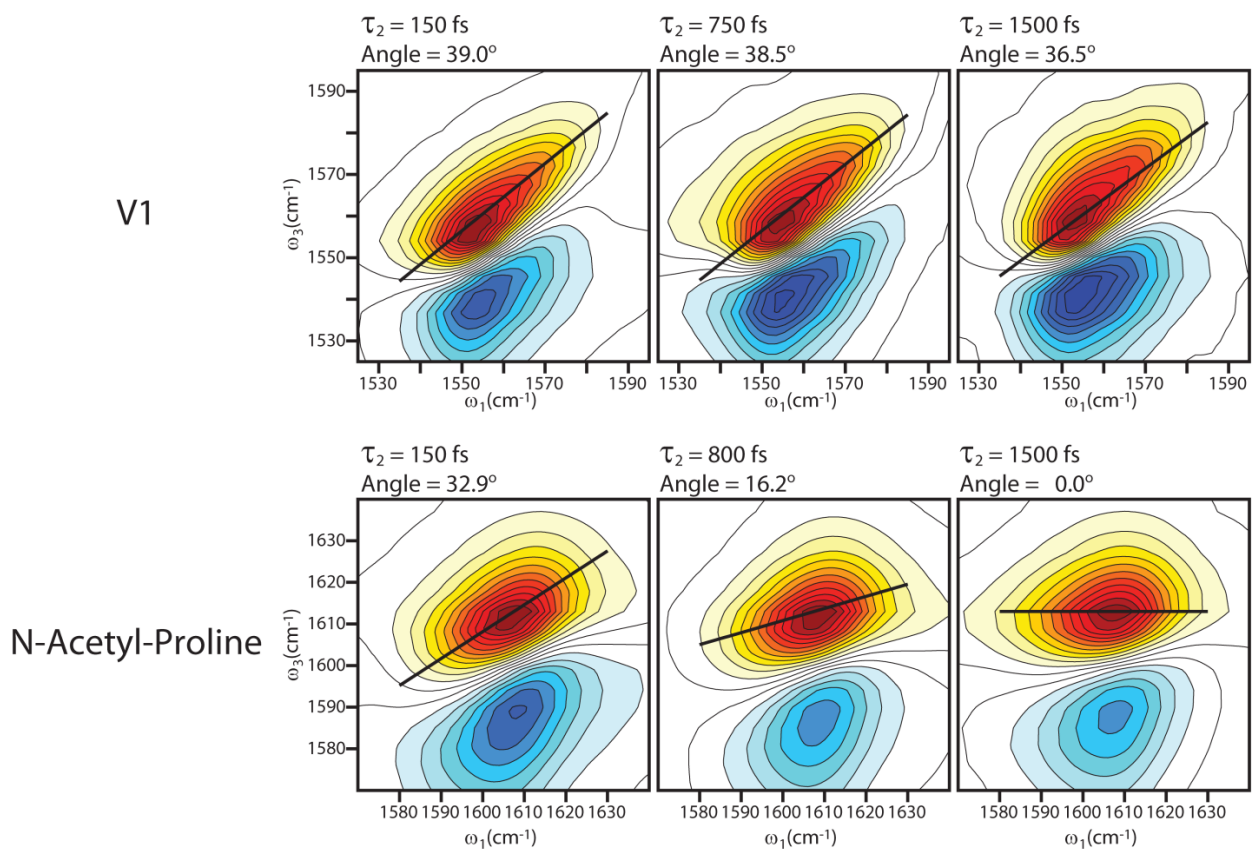

**Figure 6:** Waiting time 2D IR data for non-desalted V1 and N-Acetyl-Pro-COOD collected in ZZZZ polarization at 10 °C in pH = 1.0 DCl in $D_2O$ with no additional salt added. The black lines in the figure are the result of fitting the fundamental peak maximum.

A comparison of the 2D IR waiting time spectrum for V1 and N-Acetyl-Proline-COOD provides further proof of the coexistence of 1/0 and 2/0 turns in the equilibrium ensemble of GVGn1 (Figure 6). Figure 6 shows that when the proline amide unit is not incorporated into a peptide, as is the case for N-Acetyl-Proline-COOD, it contains a more symmetric line shape that becomes homogeneous with waiting time. The free proline peak has $\Delta = 27.6$ cm$^{-1}$, which is 69% and 43% larger than the $\Delta$ for the 1/0 and 2/0 peaks, respectively. This time scale and line width are consistent with the presence of a water exposed amide unit because water exposure is known to generate broad line shapes, and the observed relaxation time for this peak is similar to the relaxation time of the OH stretch of HOD in $D_2O$.[56] In contrast, the V1 peak does not rotate as a function of waiting time. This lack of rotation combined with the asymmetry of the peak is consistent with the existence of two overlapping peaks. In this scenario, both peaks become homogeneous with waiting time. However, because they are displaced along the diagonal axis, they appear as single static inhomogeneously broadened peaks.

**Single Site Analysis: G3 and V4**

Ab initio calculations have shown that every hydrogen bond to the amide C=O generates a 20 cm$^{-1}$ redshift to the site frequency, whereas every hydrogen bond to the amide N-H generates a 10 cm$^{-1}$ redshift.[57] As a result, the observation that both V4 and G3 have similar site energies and peak rotations as functions of waiting time (Figure 7) reinforces the proposed structure from Lessing et al.[54] in which these sites on average have their N-H pointing in the same direction relative to the peptide turn.

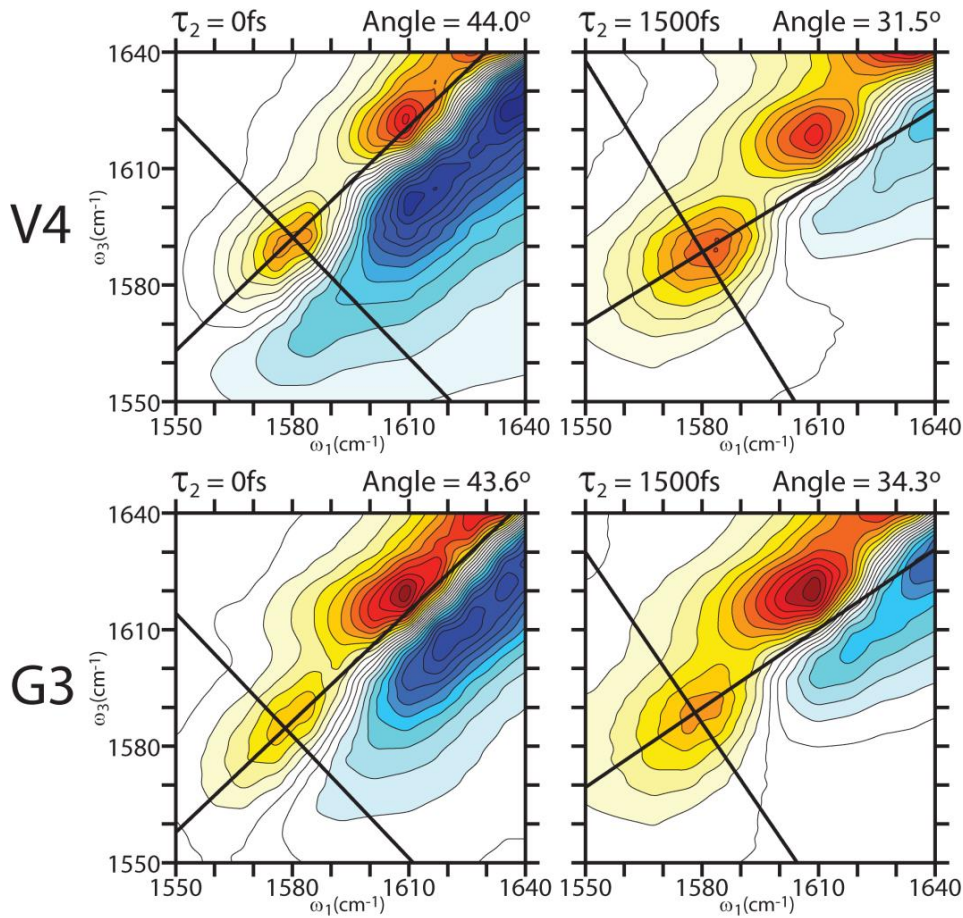

**Figure 7.** Waiting time 2D IR data for non-desalted V4 and G3 collected in ZZZZ polarization at 10 °C in pH = 1.0 DCl in $D_2O$ with no additional salt added. Each Figure has two perpendicular lines: one indicating the fit line for the fundamental peak maximum and a perpendicular line in the anti-diagonal direction provided as a guide for the eye.

This result is in contrast to the pattern of site energies and waiting time dynamics expected for a standard hairpin with a type II β turn. For a standard hairpin, these values are found to oscillate along the peptide backbone due to the alternating orientations of the amide units as observed for Trpzip2 in the folded and frayed state.[52]

Despite these similarities, there are differences between the G3 and V4 sites. First, the G3 site has a Δ of 34.6 $cm^{-1}$, which is 42% larger than the value of 24.4 $cm^{-1}$ found for V4.

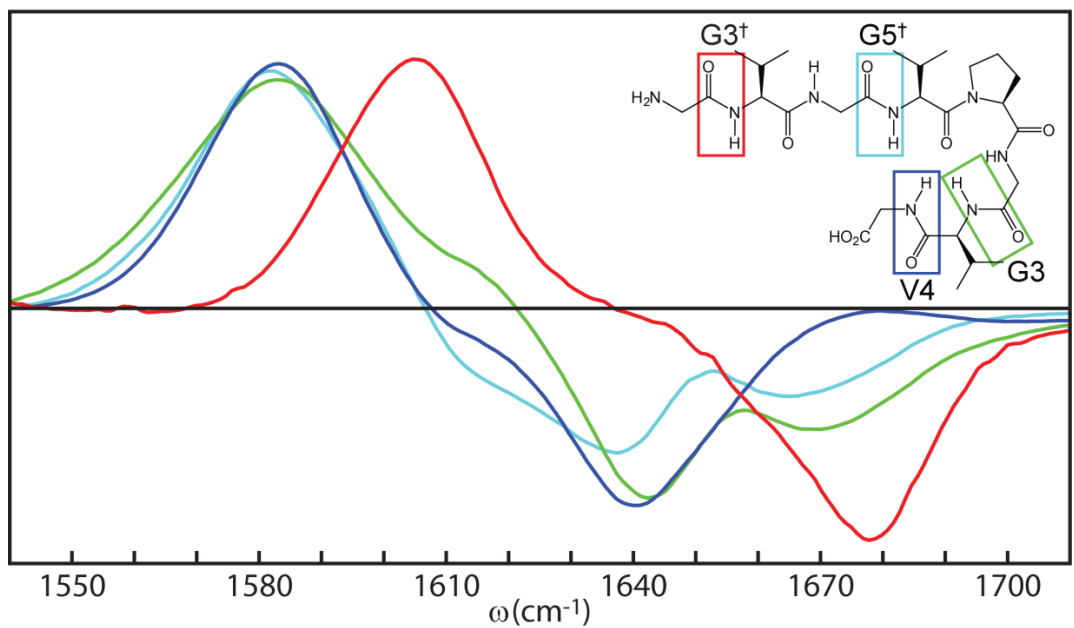

**Figure 8.** [(Isotope Label) – WT] FTIR difference spectrum for G3† (Red), G5† (Cyan), G3 (Green), and V4 (Blue), collected at 10 °C on desalted peptides in pH = 1.0 DCl in $D_2O$ with 150 mmol of $KD_2PO_4$.

Second, the G3 and V4 sites display different loss features in the FTIR difference spectrum (Figure 8). Figure 8 shows that for V4, the intensity is primarily lost from the $v_\perp$ and random coil region of the spectrum, which is consistent with a partially disordered site that participates in a conformation containing a cross strand hydrogen bond. In contrast, the G3 label displays loss features across the amide I′ spectrum, potentially demonstrating a larger diversity of possible solvent conformations for the G3 amide site. The broader loss spectrum and increased Δ in G3 are consistent with a less constrained amide site due to the absence of steric hindrance created by a side-chain moiety. Alternatively, the broad loss feature could be the result of a strong coupling between the G3 site and its neighboring amide units. In the G3 FTIR difference spectrum, there is a local maximum at ω = 1657.6 cm$^{-1}$ and two local minima at ω = 1642.2 and 1669.3 cm$^{-1}$; these values are similar to those for the random coil, $v_\perp$, and $v_\parallel$ peaks, which have center frequencies at 1660, 1630, and 1670 cm$^{-1}$. Thus, this loss pattern is also consistent with a strong coupling generating a large splitting between $v_\perp$ and $v_\parallel$ with only a small contribution to the random coil region. Finally, the FTIR difference spectrum for G3 gains intensity whereas V4 loses intensity in the region from 1610 to 1620 cm$^{-1}$, corresponding to the center of the proline peak. The observed changes could result from a change in coupling, with the G3 site gaining and the V4 site losing coupling to proline upon labeling. The G3 and V4 coupling to the V1 site will be revisited below using 2D IR spectra for analysis.

**Single Site Analysis: G5$^\dagger$**

G5$^\dagger$ has a site frequency of 1582.0 cm$^{-1}$, which is similar to the site frequency found for G3 and V4. Based on this finding, the G5$^\dagger$ site is likely to on average have a solvent exposed C=O and a N-H that is pointed towards the center of the peptide, as predicted for G3 and V4. To confirm this assertion, 2D IR waiting time measurements must first be carried out to compare the rate of nodal rotation for the G5$^\dagger$ isotope peak to the rates of G3 and V4 to determine if their solvent environments have similar dynamics. In addition, the G5$^\dagger$ FTIR loss feature is found to be similar to the G3 loss feature in that it contains intensity decreases at $v_\perp$, $v_\parallel$, and random coil frequencies. In contrast, the G5$^\dagger$ loss feature is broader and has smaller losses in the $v_\perp$ and $v_\parallel$ regions compared to G3, which reflects weaker coupling of the G5$^\dagger$ site with its neighboring amide units. Finally, the G5$^\dagger$ site is found to have a $\Gamma = 9.6$ cm$^{-1}$, giving an ellipticity of E = 0.75. This value for E is 10.7% and 8.5% smaller than the values measured for G3 and V4, respectively. The smaller ellipticity for G5$^\dagger$ may indicate that this site is more solvent exposed, and because the V4 site is on average hydrogen bonded to V1, as opposed to G5$^\dagger$, which would be the case in a standard hairpin, the G5$^\dagger$ site is unlikely to participate in peptide-peptide hydrogen bonds with sites in the turn.

## Single Site Analysis: G3†

Because the GVGn1 peptide is enriched in glycine residues, it displays a high degree of conformational flexibility. Thus, the N-terminus can bend inwards towards the turn to form peptide-peptide hydrogen bonds. Therefore, the G3† label was synthesized to explore the local environment around the N-terminus (Figure 9).

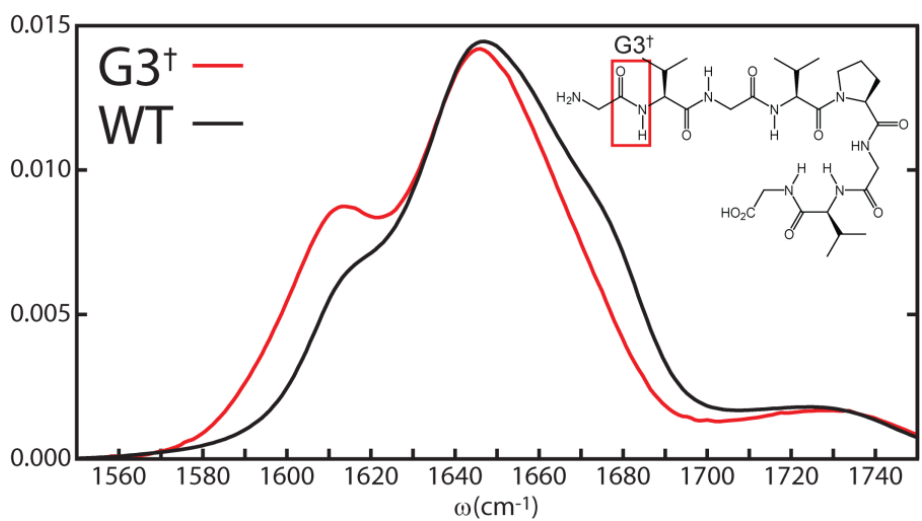

**Figure 9.** FTIR spectrum for G3† (Red) and WT (Black) collected at 10 °C on desalted peptides in pH = 1.0 DCl in D$_2$O with 150 mmol of KD$_2$PO$_4$.

The G3† site energy is located at 1604.4 cm$^{-1}$, which is blue shifted by 21.4 cm$^{-1}$ relative to the average site frequency for the G5†, G3, and V4 sites, indicating that the G3† site has fewer hydrogen bonds or a different orientation of its amide C=O relative to the turn than the G5†, G3, and V4 sites. Unlike the G5†, G3, and V4 sites, the depletion in intensity in the FTIR difference spectrum for G3† is concentrated in a frequency region to the blue of the $v_\parallel$ with a small random coil component. Here, the absence of a loss feature at the $v_\perp$ frequency might indicate that the G3† site on average does not participate in peptide-peptide hydrogen bonds. These results for G3† are similar to the findings for the S1 label of Trpzip2 presented in Smith et al.[52] This work proposed that the S1 site has a high degree of conformational mobility, which would be expected for a terminal amide group, but simultaneously has an amide C=O that is shielded from the solvent generating a high-frequency S1 peak. Given the high site frequency for G3†, it also likely has an amide C=O that is shielded from the solvent. This assignment could be explained by a conformation in which the N-terminus bends back towards the turn with its amide C=O oriented towards the V1 and V4 hydrophobic side-chains. However, modeling of the G3† site energy as a function of peptide conformation will be necessary to verify this assertion.

**Multiple Site Analysis**

In this section, the off-diagonal peak structure generated from the formation of excitonic vibrations will be explored to determine the strength of the peptide turn. The G5†G3, V1V4 and V1G3V4 peptides were synthesized to quantify the stability of the turn. By spectroscopically isolating these combinations of amide sites, it is possible to detect cross-strand couplings that are indicative of turn formation.

This approach was utilized previously in the form of the TT dual label to measure the stability of the mid-strand peptide contacts found in the Trpzip2 peptide.[52] In this example, the isotope peak of the TT double label redshifted relative to the T10 single-label isotope peak due to cross-strand coupling between the local vibrations of T3 and T10. Because the G5† and G3 site energies are nearly degenerate ($\Delta = 1.0$ cm$^{-1}$), these local vibrations can also mix to generate two new peaks whose $\Delta$ frequency splitting and cross-peak intensities are related to the coupling strength between the sites.

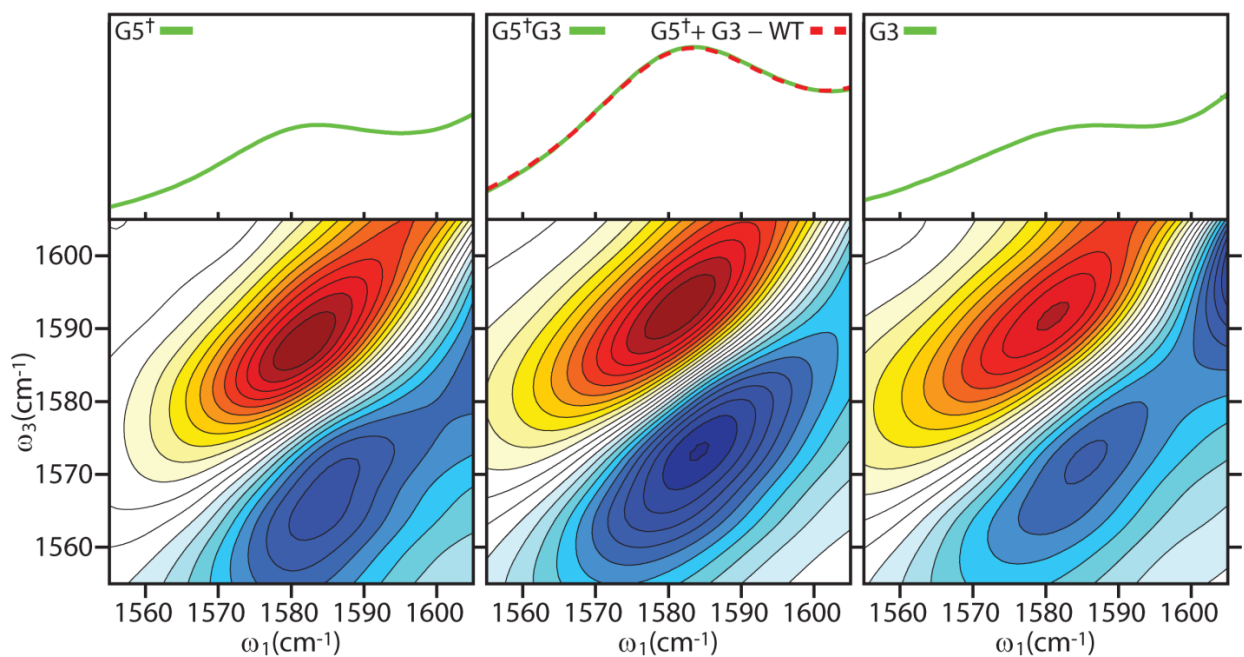

**Figure 10.** FTIR and 2D IR spectra for G5†, G3, and G5†G3 collected at 10 °C on desalted peptides in pH = 1.0 DCl in $D_2O$ with 150 mmol of $KD_2PO_4$. 2D IR spectrum were collected in ZZYY polarization at $\tau_2$ = 150 fs.

For the conditions studied here, no indications of cross-strand coupling of the G5† and G3 sites were detected, as shown in Figure 10. Visual inspection of the FTIR and 2D IR data for G5†, G3 and G5†G3 peaks indicates that the dual-labeled G5†G3 spectrum is the result of a linear combination of the single-labeled G5† and G3 spectra. In addition, the peak metrics for the G5†G3 peak ($v_{max}$ = 1581.7 cm$^{-1}$, $\Delta$ = 29.7 cm$^{-1}$, $\Gamma$ = 10.0 cm$^{-1}$, and E = 0.79) are found to be nearly identical to the average of the values for the G5† and G3 spectrum $v_{max\ average}$ = 1581.5 cm$^{-1}$, $\Delta_{average}$ = 30.1 cm$^{-1}$, $\Gamma_{average}$ = 10.0 cm$^{-1}$, and $E_{average}$ = 0.79.

Unlike the G5†G3 spectra, the G3, V1V4, and V1G3V4 spectra contain indications of cross-strand coupling. Figure 11 shows that both the V1V4 FTIR and 2D IR spectra contain features that differ from linear combinations of their corresponding single-labeled spectra. In the V1V4 FTIR spectra, these differences appear as an increase in peak intensity for the V1 2/0 peak and an increase in intensity on the blue edge of the V4 peak. In the 2D IR spectrum for V1V4 and V1G3V4, cross peaks are observed between the V4 and V1 isotope peaks (Figure 11). These cross peaks are most easily observed in the lower right corner of the spectrum as a negative ridge along $\omega_3 \sim 1544$ and $1562$ cm$^{-1}$.

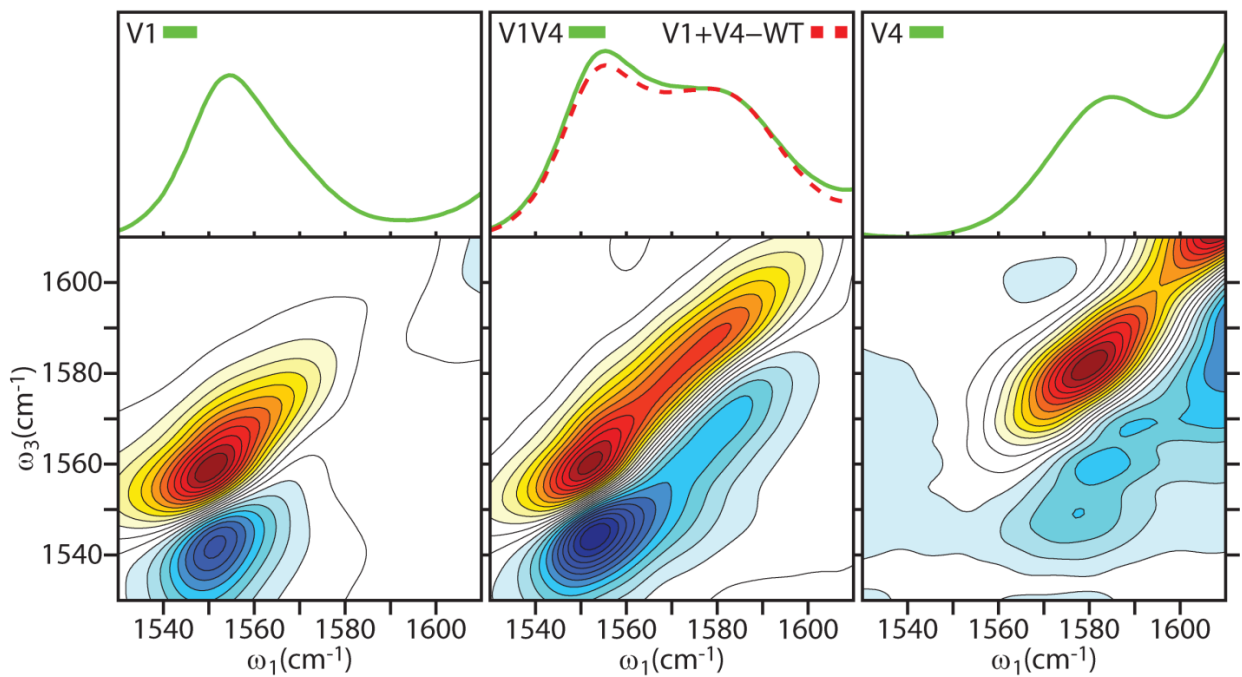

**Figure 11.** FTIR and 2D IR spectrum for V1, V4, and V1V4 collected at 10 °C on desalted peptides in pH = 1.0 DCl in D$_2$O with 150 mmol of KD$_2$PO$_4$. 2D IR spectrum were collected in ZZYY polarization at $\tau_2 = 150$ fs.

A cross peak is also observed in the G3 peptide 2D IR spectrum between the G3 isotope peak and the unlabeled proline V1 peak. This cross peak is most easily observed in the upper left corner of the spectrum as a positive ridge along $\omega_3 \sim 1624$ cm$^{-1}$. Because the G3 and V4 sites are separated from the V1 site by 1 and 2 amino acids, respectively, these cross peaks are more likely to result from a through-space as opposed to through-bond coupling. Based on the modeling results for the proline peak in Lessing et al.[54], these cross peaks are likely the result of a hydrogen bond between the G3 and V4 amide N-H moieties and the V1 amide C=O. Simultaneously, the G5$^\dagger$G3 2D IR spectrum did not indicate the existence of a longer range coupling across the turn. As a result, it can be concluded that the GVGn1 forms a closed but loose β-turn with V1 to G3 and V1 to V4 hydrogen bond contacts. To further explore these findings, it would be prudent to synthesize the $^{13}CO$ analogs of the V4 and G3 peptides and the $^{13}CO$-labeled G3V4 peptide. By using a $^{13}CO$ instead of a $^{13}C^{18}O$ label, the V4 and G3 site frequencies would be nearly degenerate with the unlabeled V1 site. In the case of the proposed $^{13}CO$-labeled G3 peptide, the frequency splitting between the labeled G3 site and the unlabeled proline site would be $\Delta\omega_{max} = 3.7$ cm$^{-1}$. As a result, this peptide should display a dramatically redshifted isotope peak akin to what was observed for the Trpzip2 TT label if the peptide has a G3 to V1 hydrogen bond.

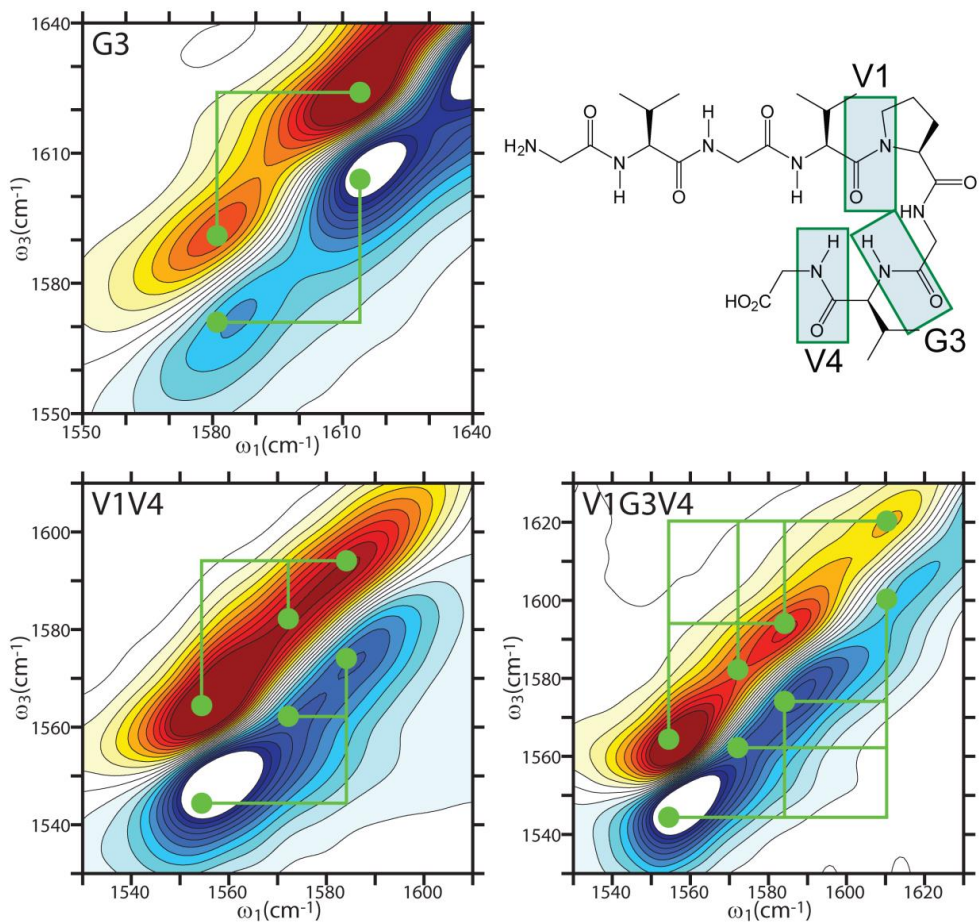

**Figure 12.** 2D IR spectra for G3, V1V4, and V1G3V4 collected at 10 °C on desalted peptides in pH = 1.0 DCl in $D_2O$ with 150 mmol of $KD_2PO_4$. 2D IR spectrum were collected in ZZYY polarization at $\tau_2$ = 150 fs. The green circles indicate the locations of isotope peaks and the corresponding green lines are provided to indicate the locations of cross peaks. In the G3 spectrum, the green dots are placed over the peak maximum of the G3 and unlabeled V1 peaks with $\omega_{max}$ = 1581.0 and 1614.0 cm$^{-1}$, respectively. In the V1V4 spectrum, the green dots are placed over the labeled V1 2/0 and 1/0 peaks with $\omega$ = 1554.4 and 1572.2 cm$^{-1}$ along with a dot over the V4 peak maximum at $\omega_{max}$ = 1584.1 cm$^{-1}$. The V1G3V4 peptide has green dots over the V1 2/0 and 1/0 peaks with $\omega$ = 1554.4 and 1572.2 cm$^{-1}$ along with a dot over the V4 and G3 peak maxima at $\omega_{max}$ = 1584.1 and 1610.3 cm$^{-1}$, respectively.

## Temperature and Salt Dependence

As discussed previously, tropoelastin and long ELPs undergo a phase transition from a structure that is soluble at low temperature to a 50-60% hydrated coacervate phase above the ITT.[23, 34, 38] Nevertheless, the changes to the amide I' spectrum of the small ELPs studied here are less dramatic (Figure 13).

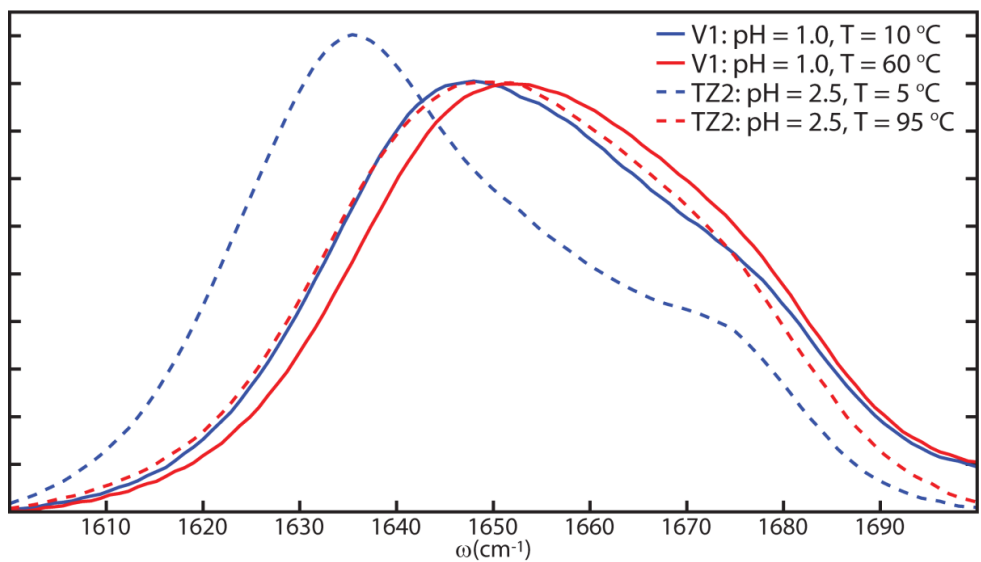

**Figure 13:** FTIR spectra for non-desalted Trpzip2 in pH = 2.5 DCl in D$_2$O with no salt added and non-desalted V1 in pH = 1.0 DCl in D$_2$O with no salt.

Figure 13 is a comparison of FTIR spectra taken at the end points of the Trpzip2 and V1 melting and "folding" transitions, respectively. As discussed in the work of Smith et al.[52], at low temperatures, the Trpzip2 peptide displays a large $v_\perp$ contribution that decreases upon heating as the cross-strand beta-sheet contacts become thermally disordered. In contrast, the high-temperature spectrum has a small $v_\perp$ contribution and is dominated by $v_\parallel$ and random coil vibrations. Interestingly, the high-temperature Trpzip2 spectrum is qualitatively similar to the V1 spectrum for the putatively "unfolded" 10 °C and "folded" 60 °C states. This comparison demonstrates both that the GVGn1 peptide exists largely in frayed and disordered conformations at all temperatures and that the changes in secondary structure for the GVGn1 peptide across the "folding" transition are small.

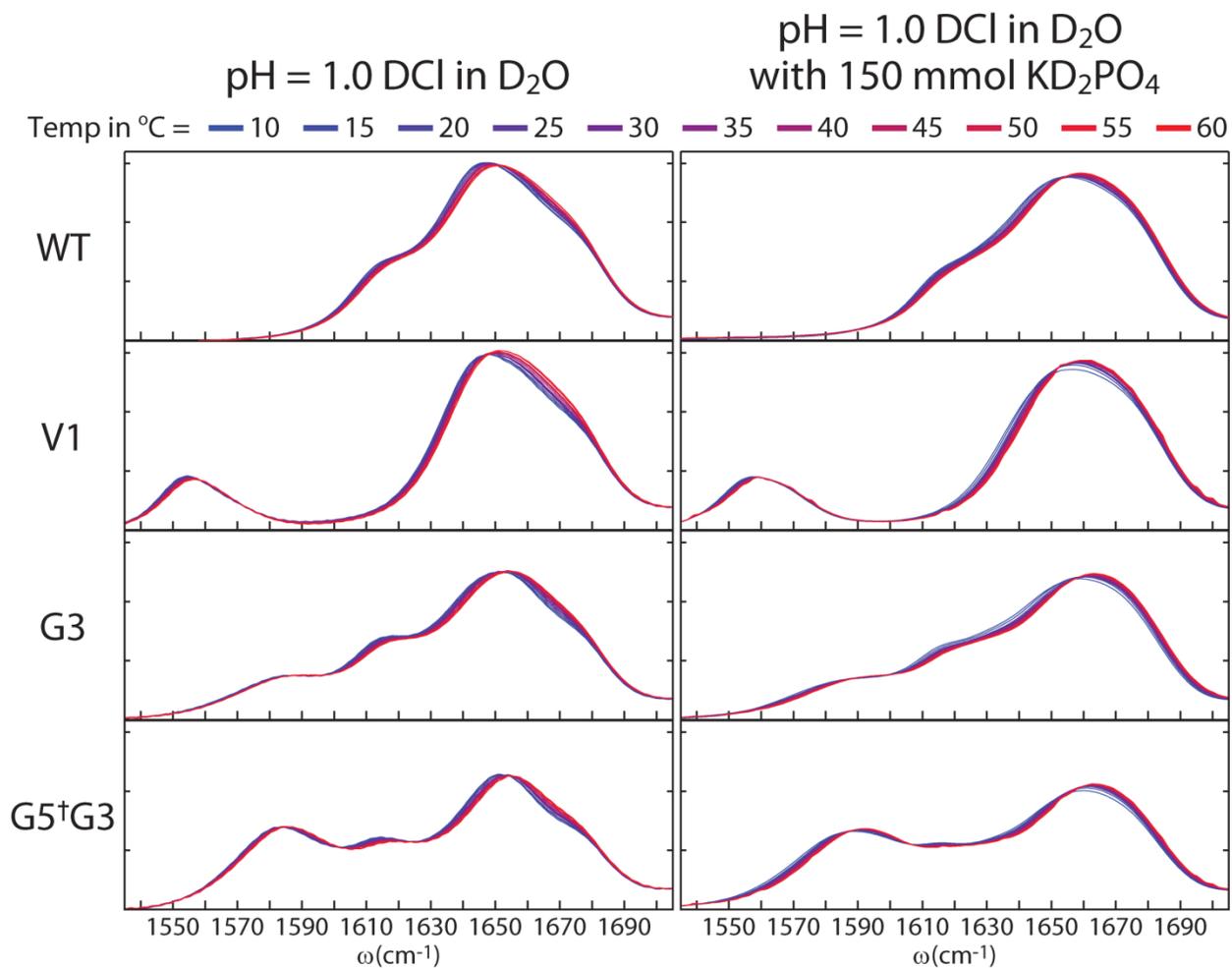

**Figure 14.** FTIR spectra for non-desalted WT, V1, G3, and G5†G3 collected as a function of temperature from 10-60 °C in 5 °C steps. The spectra in the left column were prepared in pH = 1.0 DCl in D$_2$O, and those in the right column were prepared in pH = 1.0 DCl in D$_2$O with 150 mmol of KD$_2$PO$_4$ added.

Only small changes are observed in the GVGn1 amide I′ spectrum upon increases in temperature and the addition of salt. Figure 14 shows temperature-dependent FTIR spectra for WT, V1, G3, and G5†G3 in solutions containing DCl in $D_2O$ and in solutions containing DCl and $KD_2PO_4$ in $D_2O$. Here, $KD_2PO_4$ and DCl were added to the solution to create the $D_4PO_4^+$, $K^+$, and $Cl^-$ ions to induce protein aggregation.[58-59] This experiment indicates that no substantial changes are observed for any peptide as a function of temperature regardless of salt concentration. Instead, a small blueshift is observed for the entire amide I′ line shape upon heating. Because this blueshift is small and continuous with temperature, it is attributed to nonspecific thermally induced disordering of the sample and is unlikely to be the result of a structural phase transition. Similarly, the addition of salt at constant temperature generates a blueshift in the amide I′ spectrum (Figure 15). However, in contrast to the temperature dependent data, the addition of salt to the sample also generates a subtle change in amide I′ line shape. This change is most pronounced in the region from 1660 to 1690 $cm^{-1}$ where an increase in intensity is observed as a function of salt addition.

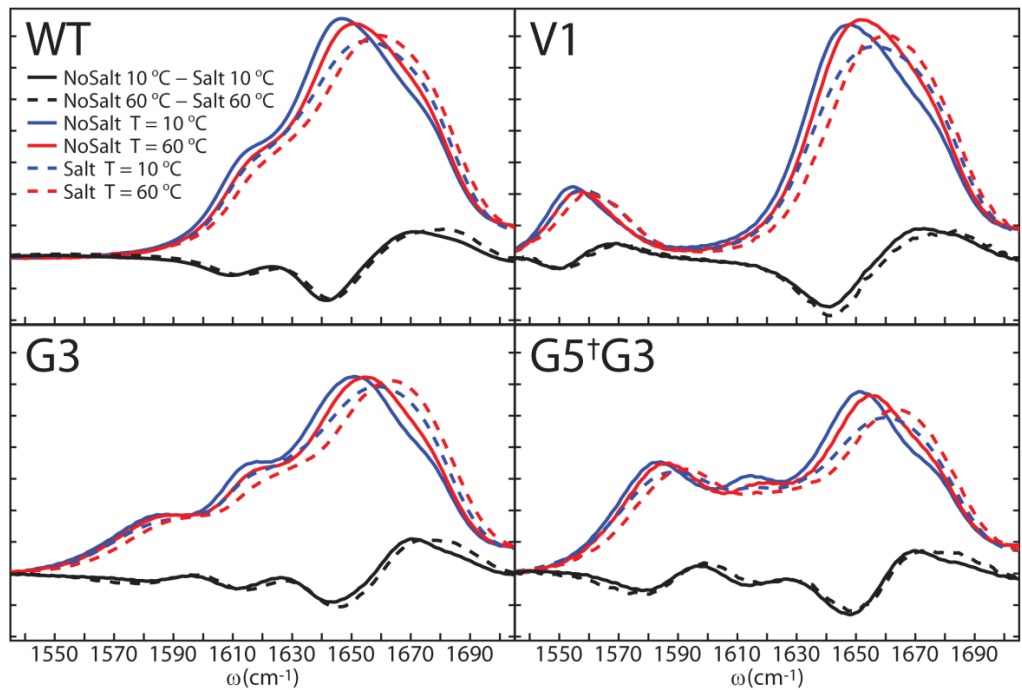

**Figure 15.** FTIR spectra for non-desalted WT, V1, G3, and G5†G3 collected at T = 10 and 60 °C. Data collected in pH = 1.0 DCl in $D_2O$ are displayed with red and blue solid lines, and data collected in pH = 1.0 DCl in $D_2O$ with 150 mmol of $KD_2PO_4$ added are displayed with red and blue dashed lines. The FTIR difference spectra (No salt – Salt) at constant temperature are displayed in black.

To quantify the effect of salt addition on turn conformation, the V1 isotope peak was fit with 2 Gaussians to determine the degree of conformational change upon heating and salt addition (Figure 16).

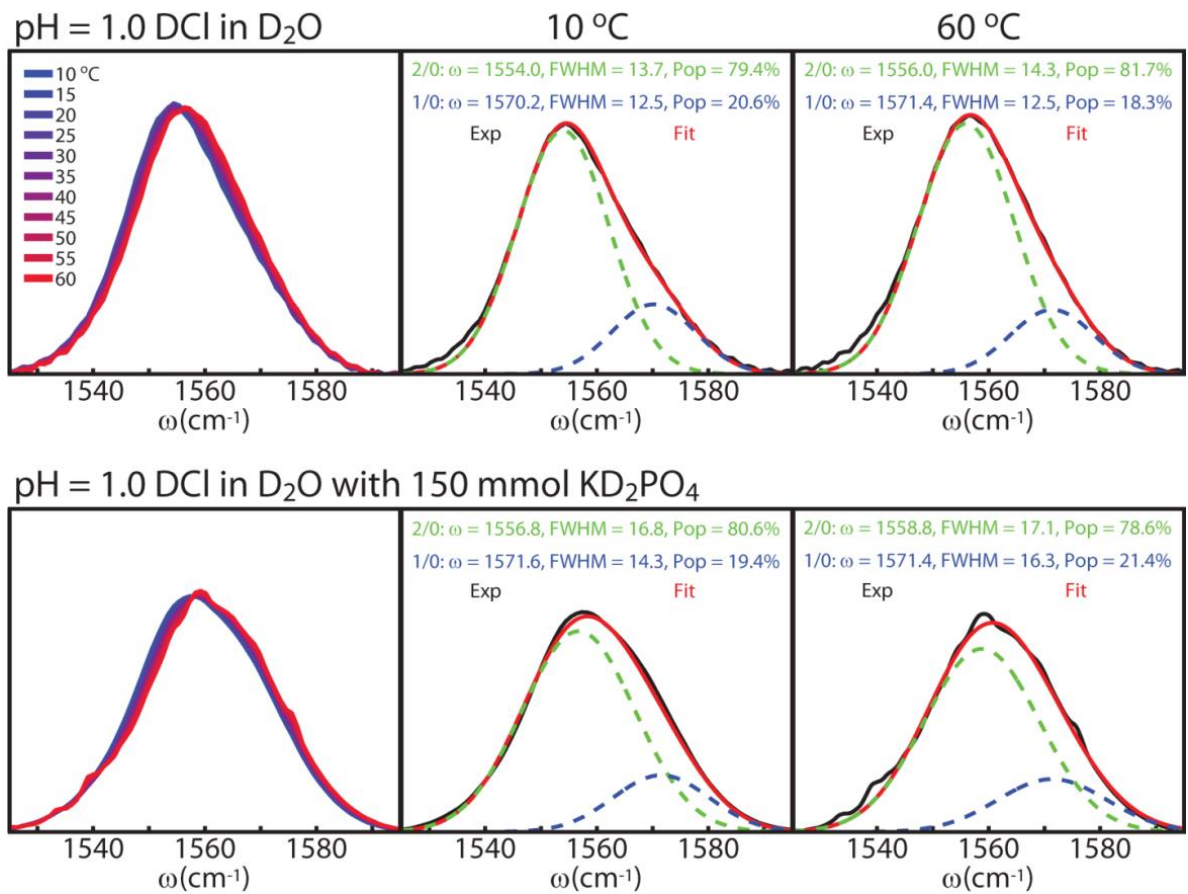

**Figure 16.** FTIR spectra collected on the non-desalted V1 peptide. (Left) Temperature dependent FTIR spectra collected from T = 10-60 °C in 5 °C steps, (Center) fit to the 10 °C spectrum, and (Right) fit to the 60 °C spectrum.

For the V1 peptide in pH = 1.0 DCl in $D_2O$ with no additional salt, the V1 peak at 10 °C is found to have 2/0 and 1/0 peaks centered at 1554.0 and 1570.2 $cm^{-1}$ with populations of 79.4% and 20.6% and FWHM of 13.7 and 12.5 $cm^{-1}$, respectively. Upon heating to 60 °C, the 2/0 and 1/0 peaks shift by $\Delta\omega$ = +2.0 and +1.2 $cm^{-1}$ with population changes of $\Delta P$ = +2.3% and -2.3% and $\Delta FWHM$ = +0.6 and 0.0 $cm^{-1}$, respectively. Similar temperature dependent results are found for the V1 peptide in pH = 1.0 DCl in $D_2O$ with 150 mmol $KD_2PO_4$. Upon heating in the presents of salt, the 2/0 and 1/0 peaks blueshift by $\Delta\omega$ = +2.0 and -0.2 $cm^{-1}$ with population changes of $\Delta P$ = -2.0 % and +2.0 % and $\Delta FWHM$ = +0.3 and +2.0 $cm^{-1}$, respectively. Small but slightly more substantial changes are observed between the salt and no salt data. Upon the addition of salt at 10 °C, the 2/0 and 1/0 peaks change by Δω = +2.8 and +1.4 cm$^{-1}$, ΔFWHM = +3.1 and +1.8 cm$^{-1}$, ΔP = +1.2%, and -1.2%, respectively. At 60 °C, the 2/0 and 1/0 peaks change by Δω = 2.8 and 0.0 cm$^{-1}$, ΔFWHM = 2.8 and 3.8 cm$^{-1}$, ΔP = -3.1% and +3.1%, respectively. The observation that the addition of salt generates a larger peak shift then an increase in temperature is also observed for the COOD resonance (Figure 17).

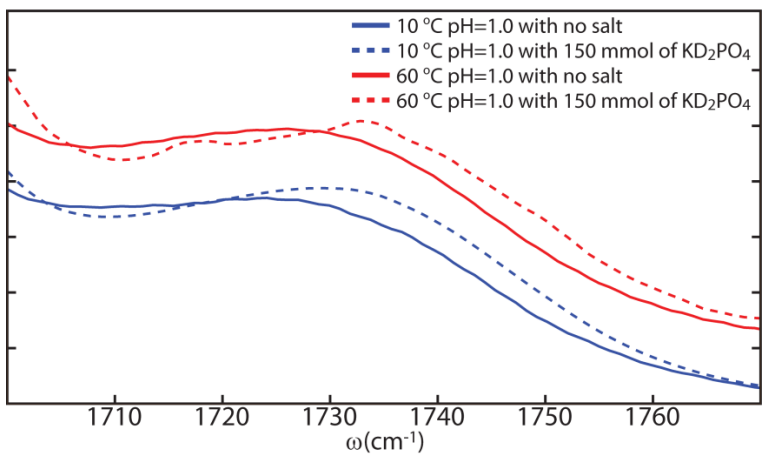

**Figure 17.** FTIR spectra collected on the non-desalted V1 peptide plotted for the COOD region.

The COOD resonance is found to blueshift by 2.7 cm$^{-1}$ upon heating from 10 °C to 60 °C and by 4.0 cm$^{-1}$ when dissolved in the 150 mmol of KD$_2$PO$_4$ solution. However, as was the case for the amide I′ resonance, these changes are continuous as a function of temperature and salt concentration, and therefore, they are unlikely to be the result of a structural phase transition. These observations indicate that the V1 and COOD peaks blueshift upon heating or the addition of salt, while the underlying 2/0 and 1/0 turn populations are unaffected to within the error of the measurements performed.

The V1 2D IR spectrum provides clearer evidence that increasing the concentration of charged species generates a larger change in the amide I′ spectra than heating 2D IR (Figure 18).

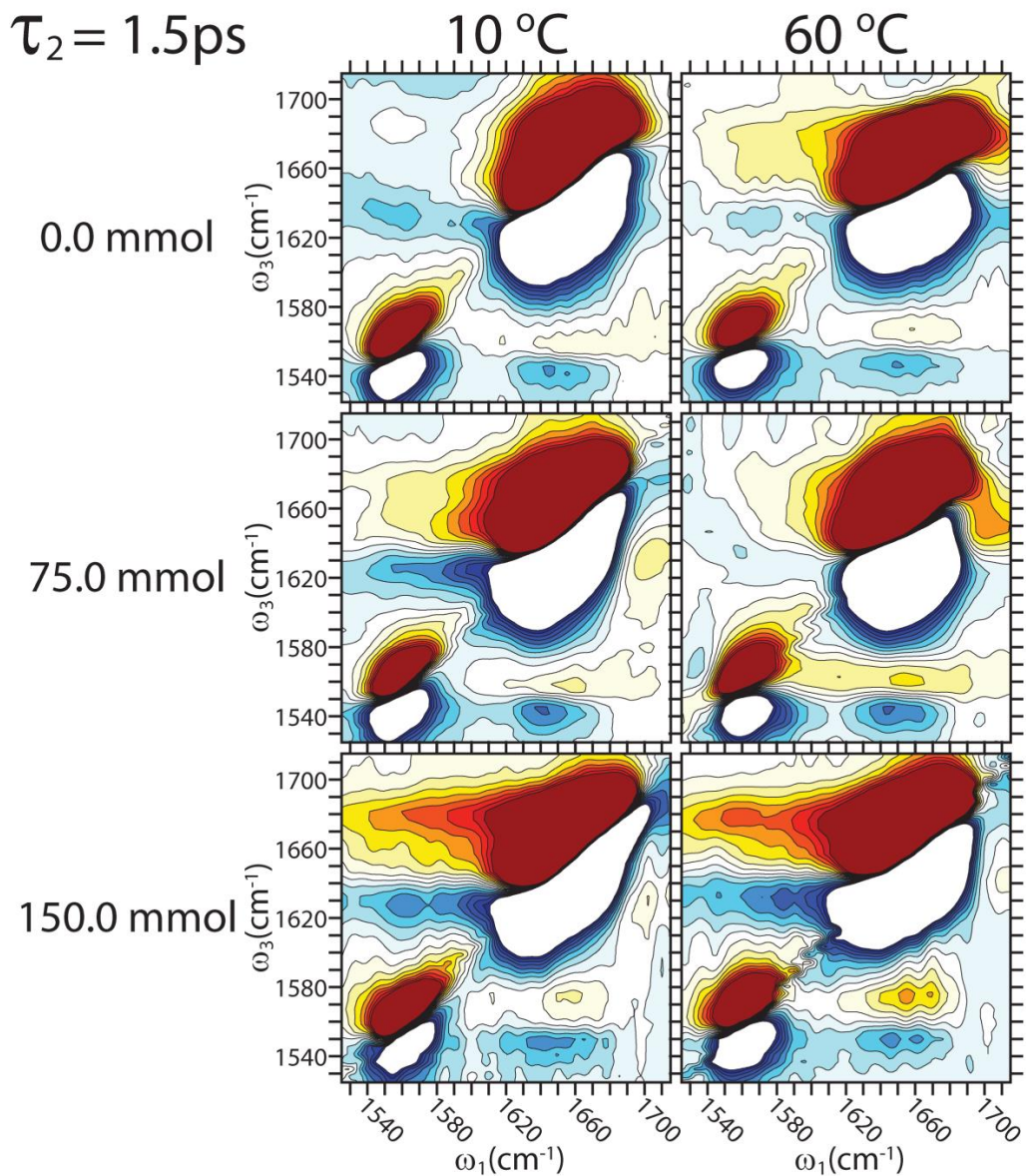

**Figure 18.** 2D IR spectra of the non-desalted V1 peptide as a function of temperature and salt concentration. All spectra were collected in a ZZZZ polarization geometry with $\tau_2 = 1.5$ ps.

In this figure, the V1 $\tau_2 = 1.5$ ps amide I′ 2D IR spectrum is given as both a function of salt concentration and temperature. In contrast to the FTIR spectra, the 2D IR spectra show pronounced changes as a function of salt concentration. Upon salt addition, the amide I′ line shape changes in the region from 1660 to 1690 cm$^{-1}$. In the 2D IR spectra, this on-diagonal line shape change occurs in conjunction with the growth of a cross peak between the 1660 to 1690 cm$^{-1}$ region and the V1 isotope peak. This cross peak is found to grow asymmetrically as a function of salt addition with a larger contribution on the blue side centered at $\omega_3 = 1678.0$ cm$^{-1}$. Because the G3$^†$ FTIR difference spectrum shows a pronounced loss feature centered at 1677.8 cm$^{-1}$, the cross peak could be the result of the N-terminus interacting with the V1 site, lending support to the previous proposal that the G3$^†$ bends inward towards the turn. Further experiments and modeling will be required to generate a structure-frequency correlation for the G3$^†$ site to confirm the existence of the proposed structure. Nevertheless, the presence of a cross peak demonstrates the existence of through-space coupling, causing us to speculate that salt addition induces structuring or a hydrophobic collapse.

**Size-dependent FTIR**

Increasing the size of the peptide is found to increase the redshift of the proline peak. This observation is illustrated in Figures 15 and 16, which show the temperature-dependent FTIR spectra and singular value decomposition (SVD) analysis for GVG(VPGVG)$_{1-6}$ and (GVGVP)$_{251}$. Figure 19 shows that the FTIR spectra for all peptides have a similar line shape with an increase in the redshift of the proline resonance as a function of increasing chain length at 5.0 °C. This redshift diminishes upon heating giving a similar proline peak shift for all peptides at 65°C. Interestingly, the first and second SVD components for all samples retain similar peak features that differ only in intensity. As a result, our understanding of the spectra of the smaller ELPs may provide the requisite insight to assign the spectrum of larger ELPs.

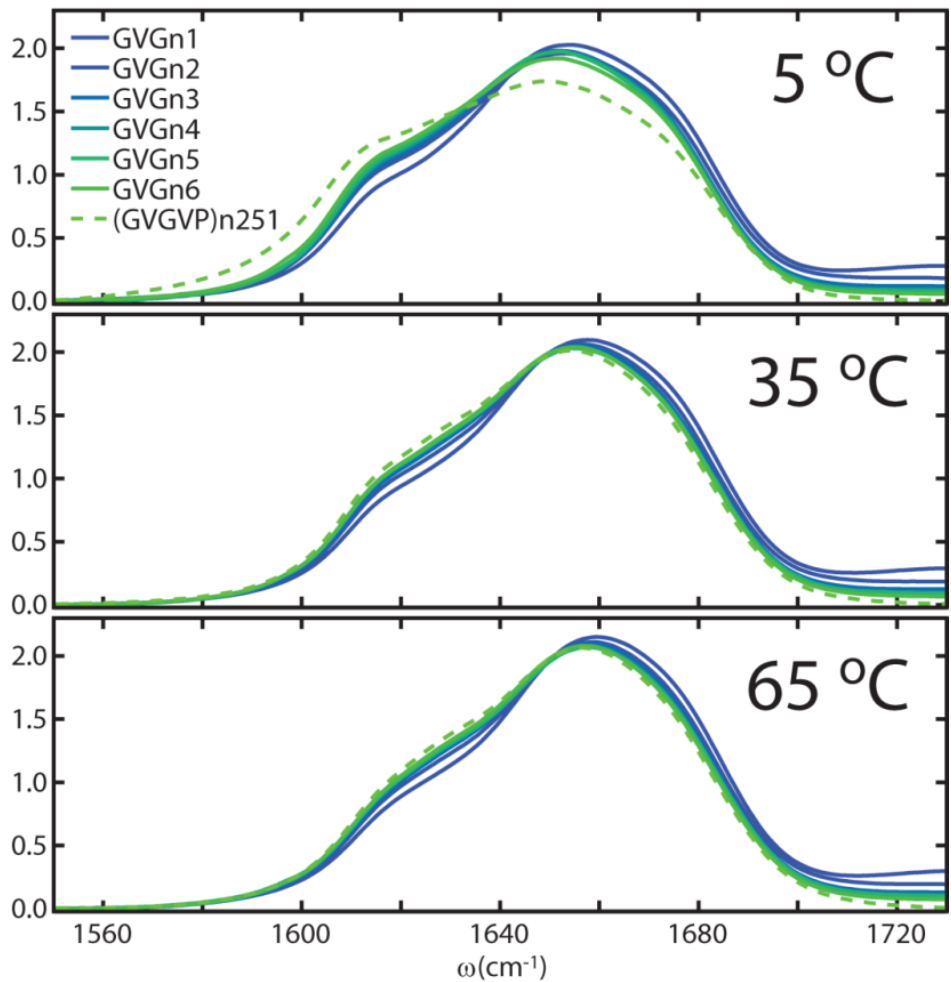

**Figure 19.** Temperature dependent FTIR spectra for non-desalted GVG(VPGVG)n where n = 1-6 and the (GVGVP)$_{251}$ polymer in pH = 1.0 DCl in D$_2$O with 150 mmol of KD$_2$PO$_4$.

Figure 20 shows that (GVGVP)$_{251}$ has an abrupt change in its first and second component SVD melting curves centered at 17 °C. In addition, the FTIR data for (GVGVP)$_{251}$ shown in Figure 21 indicate that the majority of the temperature-dependent redshift of the proline peak occurs over the temperature range 10-20 °C. Because this sample was found to transition from a low temperature state that does not scatter light to one that scatters light above 17 °C, the observed temperature-dependent changes in the IR spectrum are believed to be the result of the ITT.

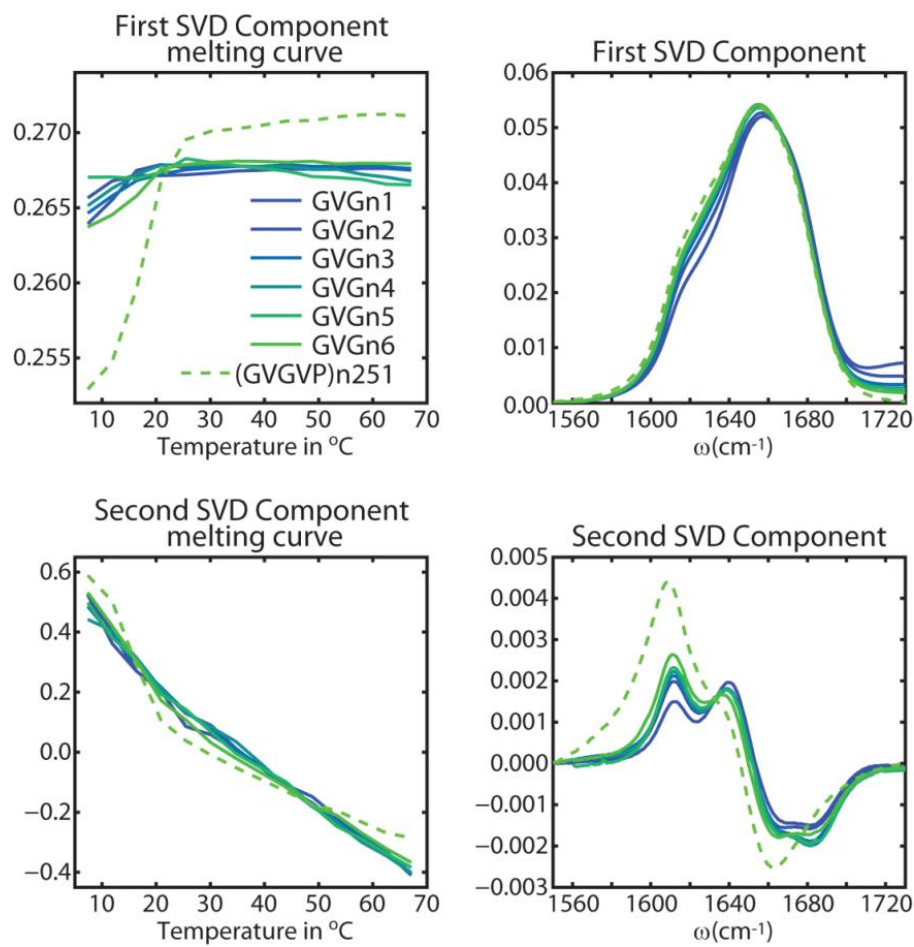

**Figure 20.** First and second SVD component spectrum and corresponding melting curves for GVG(VPGVG)$_n$ where n = 1-6 and the (GVGVP)$_{251}$ polymer in pH = 1.0 DCl in D$_2$O with 150 mmol of KD$_2$PO$_4$.

The second component spectra for GVG(VPGVG)$_n$, where n = 1-6, were found to contain local maxima at ω ~ 1612 and 1640 cm$^{-1}$ and local minima at ω ~ 1665 and 1681 cm$^{-1}$. The local maxima represent a blueshift or loss of peak intensity from the proline 2/0 peak and $v_\perp$ antiparallel β-sheet modes with temperature, and the local minima represent a blueshift or gain of peak intensity for the $v_\parallel$ antiparallel β-sheet mode and an unassigned mode at ω ~ 1681 cm-1 with temperature. To characterize the conformational changes that occur for the peptide across the phase transition the (GVGVP)$_{251}$, FTIR spectra were compared to the simulated 2/0, 1/0, and 0/0 FTIR spectra of GVGn1 from Lessing et al.[54]

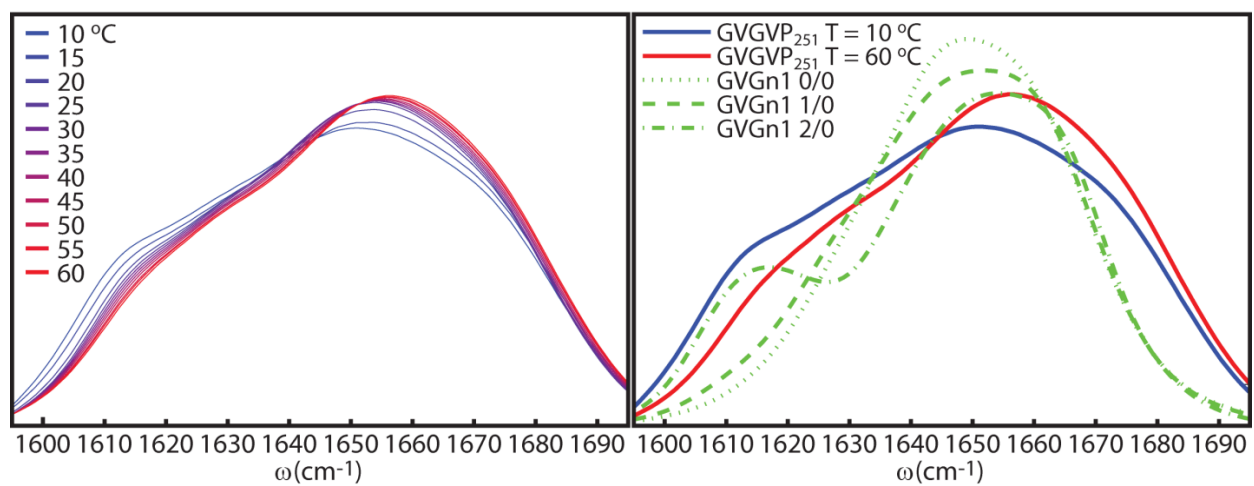

**Figure 21.** (Left) Temperature-dependent FTIR spectra for (GVGVP)$_{251}$ in pH = 1.0 DCl in D$_2$O with 150 mmol of KD$_2$PO$_4$. (Right) A comparison between the GVGVP)$_{251}$ FTIR spectrum with the simulated 2/0, 1/0, and 0/0 spectra for the GVGn1 peptide.

Figure 21 shows that the proline peak in the FTIR spectra for (GVGVP)$_{251}$ at 10 °C (proline peak maximum at $\omega = 1608.3$ cm$^{-1}$) most closely resembles the simulated GVGn1 2/0 spectrum (proline peak maximum at $\omega = 1616.9$ cm$^{-1}$). Because the proline resonance acts as a local probe of the turn structure, the redshifted proline peak for the (GVGVP)$_{251}$ polymer might have the same interpretation as in the case of the GVGn1 peptide. If this assumption is correct, the above temperature-dependent FTIR spectra would suggest a new explanation for elastin's ITT in which the peptide transitions from a low-temperature state containing primarily 2/0 turns that diminishes upon heating as the population of 1/0 turns increases. The equation from Lessing et al.,[54] which can be used to determine the number of peptide-peptide hydrogen bonds to proline (i.e., $\omega_p$(cm$^{-1}$) $\approx -13.8 \cdot n_{pp} + 1642$), gives $n_{pp} = 2.4$. This unlikely prediction may potentially reflect the fact that the redshift of the proline site in (GVGVP)$_{251}$ may originate from more than just the formation of the 2/0 turn.

The differences between the structure of GVGn1 and (GVGVP)$_{251}$ are not limited to the relative population of 2/0 and 1/0 turns. Figure 22 shows a comparison of the 2D IR ZZYY spectra of GVGn1 and (GVGVP)$_{251}$ at T = 10 and 15 °C, respectively. In addition to the broader, more redshifted proline peak for (GVGVP)$_{251}$, we observe the formation of a cross peak between the $v_\perp$ and $v_\parallel$ antiparallel β-sheet modes. This indicates that the polymer, in its low temperature state, is more structured than the GVGn1 peptide under similar conditions. One potential explanation is that the 1/0 turn in (GVGVP)$_{251}$ may contain a population of type II β-turns with both a V1 to G3 and a G5† to V4 cross strand hydrogen bond generating a unit of β-sheet structure. To assign the mechanism of the ITT in (GVGVP)$_{251}$, this β-sheet structure first has to be identified, and then, the ΔG associated with the changes in this structure relative to the ΔG associated with the changes in the turn conformation, the main-chain hydration and the side-chain hydration during the ITT have to be determined.

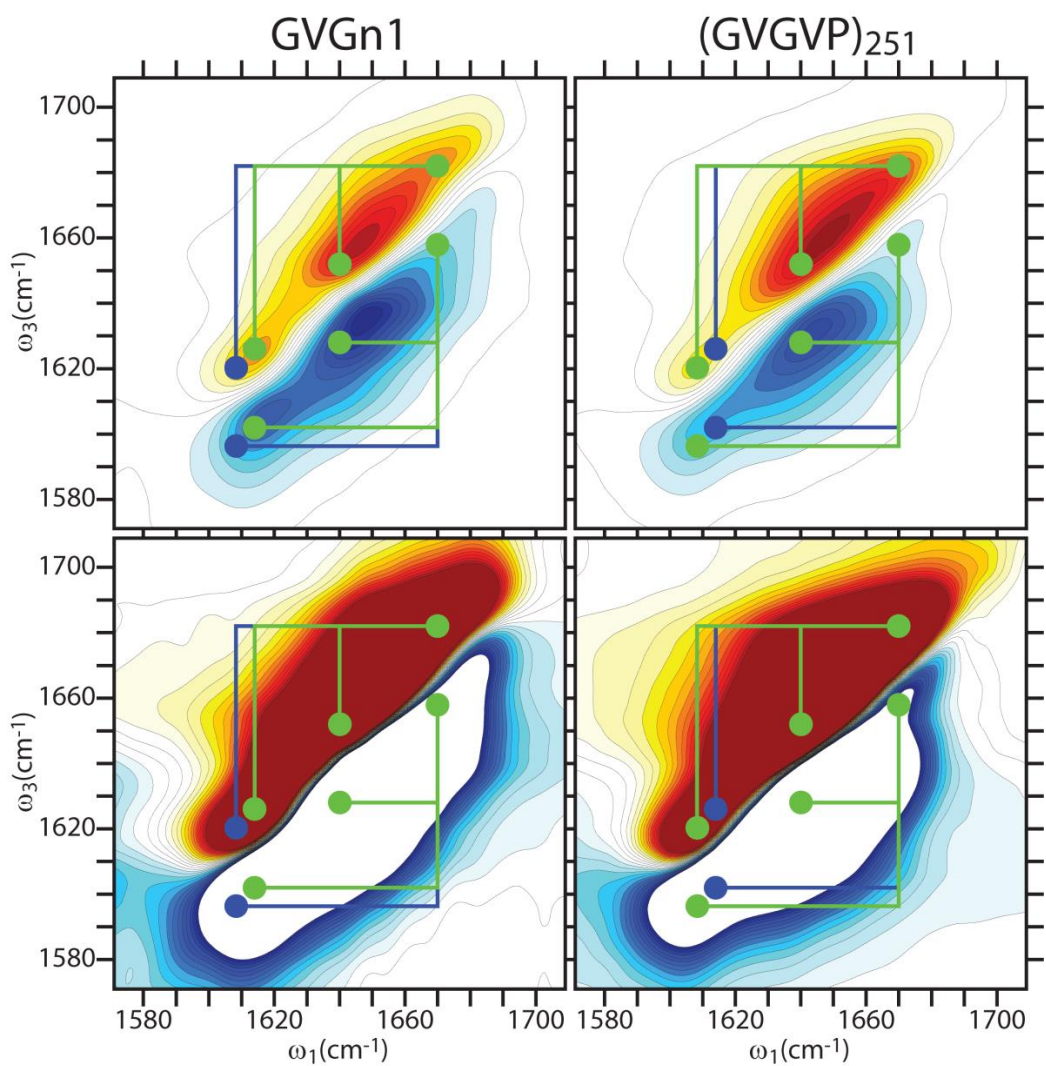

**Figure 22.** ZZYY 2D IR spectra for the GVGn1 (Left column) and (GVGVP)$_{251}$ (Right column) peptides. Desalted GVGn1 was collected at T = 10 °C and $\tau_2$ = 150 fs in pH = 1.0 DCl in D$_2$O with 150 mmol of KD$_2$PO$_4$:K$_2$DPO$_4$ (1:1). (GVGVP)$_{251}$ was collected at T = 15 °C and $\tau_2$ = 150 fs in pH = 1.0 DCl in D$_2$O with 150 mmol of KD$_2$PO$_4$. Circles are placed to mark the locations of the antiparallel β-sheet modes ($v_\perp$ = 1640 and $v_\parallel$ = 1670 cm$^{-1}$) as well as the locations of the GVGn1 and (GVGVP)$_{251}$ proline peak maxima at $\omega_{max}$ = 1614 and 1608.3 cm$^{-1}$, respectively.

# 4 Conclusions

The GVGn1 peptide on average contains a high population of closed 2/0 and 1/0 turns for all temperatures and salt concentrations investigated. The V1 C=O site is found to be buried in the interior of the turn, forming transient hydrogen bonds with the N-H of the G3 and V4 amide sites. The hydrogen bonds are observed in the 2D IR spectra of G3, V1V4, and V1G3V4 as cross peaks between their corresponding isotope peaks. The transient nature of these cross-strand hydrogen bond contacts are reflected in the G5†G3 spectra, which shows no indication of coupling between the G5† and G3 amide sites. The G5† site is unlikely to contribute a peptide-peptide hydrogen bond to the turn; instead, it is partially solvent exposed due to its location on the periphery of the peptide turn. This solvent exposure was found to decrease for the N-terminus, as demonstrated by the G3† spectrum, which contains a blueshifted isotope peak. These results indicate that the GVGn1 contains a closed irregular turn with the G3 and V4 N-H oriented towards the V1 C=O and may potentially contain a highly mobile N-terminus that is buried in the hydrophobic core of the peptide turn. This turn structure is durable, showing no significant changes with temperature or salt concentration. Based on the work presented in Lessing et al.,[54] this stability is most likely the result of steric clashes between the V1 side chain with the proline pyrrolidine ring. The addition of salt generates the largest changes in the amide I′ spectrum and a possible structuring of the N-terminus through the formation of a β-strand or interaction with the V1 site in the turn. The addition of salt generates the most pronounced changes in the amide I′ band and the off-diagonal cross peak region while not generating significant shifts in the isotope peak frequencies. As a result, the driving force for the phase transition for VPGVG sequences is likely due to a hydrophobic effect involving the aggregation of Val side chains to form a hydrophobic core, and as a result, the phase transition does not project significantly onto the amide I′ spectrum.

# 5 Experimental Details

All of the elastin-like peptides presented in this study, with the exception of the (GVGVP)$_{251}$, were prepared using Fmoc-based solid phase peptide synthesis (SPPS) on Gly-Wang resin. The Gly-Wang resin and the unlabeled Fmoc amino acids used to make the elastin-like peptides were purchased from Novabiochem. The resulting peptide samples from the SPPS procedure were purified by reverse phase HPLC using a two-phase buffer gradient: (buffer A) 0.1% TFA in H$_2$O and (buffer B) 80% acetonitrile, 0.085% TFA in H$_2$O and were desalted where indicated. These peptides were subsequently 3x dissolved in an HCl solution and lyophilized to remove TFA from the sample.[60,61] The Trpzip2 peptide was prepared according to the procedure presented in the work of Smith et al.[52] The N-acetyl-LPro-COOH was purchased from Sigma-Aldrich and was used without further purification. The recombinantly expressed (GVGVP)$_{251}$ was obtained from Dan W. Urry of Bioelastics Research Ltd. and was used without further purification. Isotope-labeled Fmoc amino acids were purchased as 99% enriched 1-$^{13}$C labels from Cambridge Isotope Laboratories. To create the $^{13}$C$^{18}$O labels, the carboxyl oxygens of the 1-$^{13}$C Fmoc amino acids were labeled with $^{18}$O via acid hydrolysis in H$_2$$^{18}$O purchased from Isoflex USA. Acid hydrolysis was conducted by first dissolving the desired Fmoc amino acid in a mixed solvent system of $H_2^{18}O$ and 1,4-dioxane in a 1:4 ratio. This solution was then acidified to pH = 1.0 using HCl generated in situ through the addition of acetyl chloride. The solution was then refluxed at 100°C for 72 hrs and lyophilized, generating a ~75% isotopically enriched sample. The labeling procedure was then repeated to achieve ~97% isotopic enrichment.

Samples for 2D IR and FTIR experiments were first dissolved in $D_2O$ and lyophilized to remove $H_2O$ and exchange acidic protons in the sample for deuterium. The dry deuterated product was then dissolved in pH = 1.0 DCl in $D_2O$ with 0, 75 or 150 mmol of $KD_2PO_4$ or pH = 1.0 DCl in $D_2O$ with 150 mmol of the mixture $K_2DPO_4$ and $KD_2PO_4$ in a 1:1 molar ratio. A sample of the 150 mmol solution of the $K_2DPO_4$ and $KD_2PO_4$ mixture was analyzed by Northeast Environmental Laboratory, Inc. (Danvers, MA), which measured the concentration of the potassium, chloride and orthophosphate ions to be $K^+$ = 207 mmol, $Cl^-$ = 256 mmol, and $PO_4^{3-}$ = 130 mmol. Elastin-like peptide samples were prepared at concentrations from 15.0 to 50.0 mg/ml with no spectral changes observed over this range. Next, samples were placed between two 1-mm-thick $CaF_2$ windows with a 50-µm Teflon spacer generating an ~25-µL sample volume. The sample was then mounted in a temperature-controlled brass sample holder. FTIR data were collected with a Nicolet 380 FTIR spectrometer in a dry air purged environment at a resolution of 2.0 $cm^{-1}$. 2D IR spectra were collected in a dry air purged environment using 100-fs broad band mid-infrared pulses with a $\tau_1$ evolution time that was scanned to 3.1 ps for rephasing spectra and 2.5 ps for non-rephasing spectra. Data were collected in both ZZZZ and ZZYY polarization geometries at a variety of waiting times ranging from $\tau_2$ = 0.0 to 1.5 ps, as indicated.

# 6 Acknowledgments

The author would like to thank Carlos Biaz, Kevin Jones, Chunte Sam Peng and Michael Reppert for assistance with the 2D IR experiments. The author would also like to thank Jongjin Kim and Sam Chunte Peng for aid with the synthesis of the isotope labeled peptides and the size dependent peptides GVGn2 through GVGn6, respectively. The author would like to thank Dan W. Urry of Bioelastics Research for providing the (GVGVP)$_{251}$ sample. Finally, the author would also like to thank Andrei Tokmakoff for his mentorship.

# 7 Author Information

*Corresponding Author E-mail: JoshuaLessing@gmail.com

Notes: The authors declare no competing financial interest.